# Estimating the Potential Impact of Combined Race and Ethnicity Reporting on Long-Term Earnings Statistics

Kevin L. McKinney, U.S. Census Bureau[1]

John M. Abowd, Cornell University

July, 2024

Abstract

We use place of birth information from the Social Security Administration linked to earnings data from the Longitudinal Employer-Household Dynamics Program and detailed race and ethnicity data from the 2010 Census to study how long-term earnings differentials vary by place of birth for different self-identified race and ethnicity categories. We focus on foreign-born persons from countries that are heavily Hispanic and from countries in the Middle East and North Africa (MENA). We find substantial heterogeneity of long-term earnings differentials within country of birth, some of which will be difficult to detect when the reporting format changes from the current two-question version to the new single-question version because they depend on self-identifications that place the individual in two distinct categories within the single-question format, specifically, Hispanic and White or Black, and MENA and White or Black. We also study the USA-born children of these same immigrants. Long-term earnings differences for the 2nd generation also vary as a function of self-identified ethnicity and race in ways that changing to the single-question format could affect.

Keywords: Statistical Policy Directive 15, Hispanic earnings, Middle Eastern and North African earnings, Immigrant children earnings

---

[1] The opinions and conclusions expressed herein are those of the authors and not those of the U.S. Census Bureau, Cornell University, or any other research sponsor. Research presented here was authorized under project P-6000266. All results have been reviewed to ensure that no confidential information is disclosed (DRB release number CBDRB-FY23-CES-011-002). This research uses data from the Census Bureau's Longitudinal Employer-Household Dynamics Program, which was partially supported by NSF Grants SES-9978093, SES-0339191, and ITR-0427889; NIA Grant AG018854; and grants from the Sloan Foundation.

## 1. Introduction

The United States has collected race and ethnicity data on every decennial census of population since 1790. The subjugation and abuse of minorities and indigenous peoples partially enabled by those data are part of the indelible history of the nation. Since 1970, collecting information on race and ethnicity has served a different purpose: providing statistical data for the enforcement of civil rights laws passed in the 1960s including the 1965 Voting Rights Act. In support of that effort, two important principles were embodied in the regulations governing these data collections. The first principle, embodied in the 1977 Statistical Policy Directive 15 (SPD 15 hereafter), required that the respondent provide the answers to questions on race and ethnicity, what we call self-identification, ending the practice begun in 1790 of the enumerator supplying that information.[2] The second principle, embodied in the 1997 revision of SPD 15, enabled respondents to characterize their racial identities in as many of the five categories as they wished. Amendments to the 1954 Census Act that are still in force added a sixth category, Some Other Race, to the regulatory five: American Indian or Alaska Native, Asian, Black or African American, Native Hawaiian or Pacific Islander, and White.[3] In March, 2024, a third principle, embodied in the 2024 revision of SPD 15, was added: the respondent may choose from any of seven categories (the five in SPD 15 plus Hispanic or Latino and Middle Eastern or North African) without any requirement to distinguish between racial and ethnic categories. The amendments to the Census Act still require the inclusion of the category Some Other Race. We call this the "single-question" format.

The 1997 revision of SPD 15 also required that information on Hispanic or Latino ethnicity be collected separately and asked before questions about race. Research in the 2010s, culminating in the 2015 National Content Test, revealed that collecting the information on race and ethnicity in a single question that included the five 1997 categories, Hispanic or Latino, and Middle Eastern or North African (MENA) and made no distinction between "racial" and "ethnic" categories provided better quality data and virtually eliminated respondent-provided "Some Other Race" answers (Matthews *et al.* 2017). The 2024 revision of SPD 15 replaced the two-question format with a single-question format like the ones tested in 2015. The design also encourages respondents to enter additional national origin detail, which is often based on place of birth (POB).[4]

[2] See Office of Management and Budget (2023) for a history of Statistical Policy Directive 15 and the proposed changes.

[3] In official publications of the U.S. Census Bureau, the category "Some Other Race" is capitalized to distinguish it from responses like "one of the other categories," "none of the above," or "other." In this paper, we use the same convention. We always mean that the respondent selected Some Other Race in lieu of, or in addition to, any of the other choices available on the instrument. Under the 1997 SPD 15, those categories are American Indian or Alaska Native, Asian, Black or African American, Native Hawaiian or Pacific Islander, and White.

The 1997 regulatory definitions of racial and ethnic categories also included rules for categorizing individuals when write-in responses provided national origin data. These included: coding any race response indicating national origin from a Spanish-speaking country as “Hispanic or Latino” ethnicity then coding “Some Other Race” for the race category, and coding “White” for many MENA countries if the respondent did not indicate a different race. Other papers at the conference (Jensen and Jones, 2024) deal with these issues in detail. We note here only the consequences that are salient for our research. Data from recent censuses and 1997 SPD 15 compliant surveys like the American Community Survey always appear to have a respondent-provided determination of Hispanic or Latino ethnicity and a respondent-provided set of racial categories (up to six, including Some Other Race). But no respondent was permitted to identify as Hispanic without also identifying within racial categories that excluded Hispanic and related categories. Similarly, no respondent was permitted to identify as Black (or any other race category) without also indicating whether they identified as Hispanic or Latino. Under the single-question format, multicategory responses must be entirely initiated by the respondent. Our research informs comparisons of differences in outcomes based on self-identification with those based on place of birth or parent’s place of birth as recorded on the individual’s Social Security Number application form. [5]

In this paper, we use data from three distinct sources—detailed place-of-birth data from the Social Security Administration (SSA), detailed race and ethnicity responses from decennial censuses of population, and earnings data from state Unemployment Insurance (UI) systems—to try to disentangle the differences in long-term real earnings associated with self-reported racial and ethnic identities as distinct from country of origin. To do so, we consider a cohort of foreign-born adults along with a similar cohort of USA-born adults. For USA-born adults we create a sub-sample of young workers, comparing outcomes for those with USA-born or foreign-born parents. Place of birth is measured from Social Security Administration data collected when the person was issued a Social Security Number. Racial and ethnic identities are measured by the detailed direct responses from the 2000 and 2010 Censuses with all allocated and edited responses removed. We characterize a place of birth as “Hispanic” if at least 50% of persons born in that country self-identify as Hispanic. We characterize a place of birth as MENA using the geographic location of the country since there is no reliable way to measure MENA self-identification in the 1997 SPD 15 format. Using the combined data sources, we attempt to measure whether the self-identification of racial or ethnic origin of immigrants corresponds to identification based on administrative place-of-birth.

---

[4] See Marks and Jones (2020) for a detailed explanation of the race and ethnicity coding rules for the 2010 and 2020 Censuses. See Waddell *et al.* (2022) for a discussion of the implications for economic measurement. Jensen and Jones (2024) consider these issues in detail.

[5] It is important to note here that “Some Other Race” is not a social-political construct. The coding rules described in the text were intended to comply with the statutory obligation to include Some Other Race and the 1997 SPD 15 regulation to treat Hispanic ethnicity distinct from any racial self-identification when collecting and tabulating decennial census data. When bound only by the 1997 SPD 15, Census Bureau publications do not include Some Other Race. Those responses are recoded to one of the five categories listed above.

Our outcome measure is average annual real labor market earnings over the 6-year period from 2010 to 2015, which we call long-term earnings. Family structure is established from the 2000 Census. We present statistics based on the measured and regression-adjusted long-term earnings organized by place-of-birth for the main sample and by the parent's place of birth for the children sample.

Section 2 describes our data sources and limitations. Section 3 describes our statistical methods. Section 4 presents the results. Finally, Section 5 concludes.

## 2. **Data Sources and Limitations**

The empirical work in this paper is based on job-level earnings information from the Longitudinal Employer-Household Dynamics (LEHD) infrastructure files, developed and maintained by the U.S. Census Bureau.[6] In the LEHD infrastructure, a job is the statutory employment of a worker by a statutory employer as defined by the UI system in each state. Mandated reporting of UI-covered wage and salary payments between one statutory employer and one statutory employee is governed by the state's UI system. Reporting covers private employers and state and local government. LEHD also receives earnings reports from the Office of Personnel Management (OPM), which provides coverage of a large share of the federal civilian workforce.[7] There are no self-employment earnings unless the proprietor draws a salary, which is indistinguishable from other employees in this case.[8]

The LEHD program is based on a voluntary federal-state partnership. When a state becomes a member of the partnership, current as well as all available historical data for that state are ingested into the LEHD internal database. By 2004, LEHD data represent the complete universe of statutory jobs covered by the UI system in the United States.[9] We use all jobs to construct person-year level annual real (deflated by the 2010 Personal Consumption Expenditures Index) earnings files covering the period 2010-2015. In our previous research we constructed earnings measures covering the entire complete universe 12-year period (2004-2015). However, in this paper we construct two samples using a shorter 6-year period from 2010-2015. The shorter 6-year time-window is a potentially less robust measure of long-term earnings but allows us to identify a young cohort of 2nd generation workers (children of foreign-born and USA-born parents) using information from the 2000 Census of Population and Housing.

---

[6] See Abowd *et al.* (2009) for a detailed summary of the construction of the LEHD infrastructure.

[7] Unemployment Insurance for federal workers is administered through the former employee's state of residence, but the states do not collect wage reports or UI premiums from the federal government.

[8] UI earnings are reported by the employer every quarter. Although there is some variability across states, UI earnings generally include regular hourly earnings or salary, overtime, bonuses, reported tips, and sick/vacation pay. Federal workers participate in the unemployment insurance system, but their earnings are reported to LEHD by the Office of Personnel Management. Private, state, local, and federal employers are all included in our analysis.

[9] We have previously shown that by 1998 missing state data do not significantly affect measures of inequality and volatility (Abowd, McKinney, and Zhao, 2018McKinney and Abowd, 2023). However, to minimize the impact of employer non-reports (false zeros) on estimates of individual long-term average earnings, we restrict our sample to the period 2010-2015, where data from all 50 states, DC, and the federal government are available.

We use methods we have employed before (*e.g.*, Abowd *et al.*, 2018) to create a longitudinal frame of persons that covers a known population of interest and has a relatively high level of confidence that the persons in that population use a consistent person identifier across time and jobs. To that end, we use the U.S. Census Bureau's enhanced version of SSA's master Social Security Number (SSN) database (the Census Numident) to create a set of "eligible" workers each year, removing annual earnings records for ineligible workers. The first eligibility condition is that the worker's SSN appears on the Numident. We call such SSNs "active" because if they have wage and salary income in the U.S., it should appear in our data. Each year an eligible worker must also meet additional conditions: age within the range specified by our two samples, not reported dead, and active SSN. The worker also must not have more than 12 reported employers during the year; otherwise, we assume the SSN is being used by multiple persons and the annual earnings report is discarded.[10] We impose no additional annual earnings restrictions. Zero-earnings years are included as long as the worker is active in the labor market at least one quarter during the analysis period.[11]

We match the set of eligible workers active in UI covered jobs during 2010-2015 to information on reported detailed race and ethnicity from the 2010 Census of Population and Housing. For Sample 1, a worker must have complete data for date of birth, place of birth, and sex on the Census Numident, be between the ages of 25 and 55 (inclusive) in 2010, be "eligible" all 6 years as defined above, and have reported race and ethnicity on the 2010 Census.[12] These rules produce an analysis Sample 1 of 87.97 million workers.[13] Sample 2 is a subset of Sample 1

---

[10] Abowd, McKinney, and Zhao (2018) provide detailed evidence that our exclusion of identifiers with 12 or more employers in a year are very likely to be associated with multiple users of the same identifier.

[11] Zero-earnings years are not directly reported in the LEHD data, a zero-earnings year is inferred based on the absence of reported earnings. Zero-earnings years are a core part of the long-term average earnings analysis and restricting our sample to the complete-data time period 2010-2015 ensures that each worker has a consistent, small probability each year of a non-report.

[12] We use the detailed race and ethnicity data from the 2010 and 2000 Censuses. These include all checkboxes and write-ins that were captured in the data processing. We use only reported responses, excluding all allocations and edits.

[13] Our analysis samples only use UI earnings records with a Protected Identification Key (PIK) found on the Census Numident. Only PIKs from the Census Numident are assigned to records on the 2010 Census by the Census Bureau's Personal Identification Validation System (Wagner and Layne, 2014). An eligible worker in the LEHD data may not match to the 2010 Census if the respondent refused to take the survey (and consequently has entirely

consisting of workers born in the U.S., age 25 to 28 (inclusive) in 2010, who have at least one reported biological parent in the home on the 2000 Census. For Sample 2, we use the Census Numident place of birth information for the reported biological parent(s) as indicated on the 2000 Census household roster to determine whether the worker was a 2nd generation immigrant (child of at least one foreign-born parent) or a 3rd (or higher) generation immigrant (child of two parents born in the U.S. if two are available or child of one parent born in the U.S. if only one is available). These rules produce an analysis Sample 2 consisting of 7.196 million workers.

## 3. Statistical Methods

We create a single long-term average annual earnings measure consisting of real annual earnings, $w_i$, summed across all years and all UI-covered employers and divided by six:

$$w_i = \frac{1}{6} \sum_{t=2010}^{2015} \sum_{j} e_{ijt} \,,$$

where $e_{ijt}$ is the real earnings of individual *i* at employer *j* in year *t*.[14] Nominal earnings were converted to real using the Personal Consumption Expenditures Price Index (2010 = 100). We use $w_i$ to explore the differences in long-term average real annual earnings across sex, place of birth, ethnicity, and race categories. Our race variable consists of the following eight categories: White, Black, American Indian or Alaska Native, Asian, Native Hawaiian or Pacific Islander, Some Other Race, 2+ Races (not SOR), and 2+ Races (SOR). The notation 2+ Races (SOR) means that one of the coded race categories is Some Other Race, and 2+ Races (not SOR) means that none of the coded race categories is Some Other Race. The ethnicity variable is binary: Hispanic or Latino, Not Hispanic or Latino.

Our universe of Hispanic-origin countries was created by tabulating the percentage of individuals who self-identified as Hispanic or Latino on the 2010 Census for every country of birth in Sample 1. We classify a place of birth as Hispanic when at least 50% of the immigrants from that country self-identify as Hispanic.[15] No similar exercise works for MENA countries.

---

imputed data), the respondent took the survey but did not report some demographic data (and consequently received a PIK not in the LEHD data), or the respondent record was not assigned a PIK. In addition, an eligible worker in the LEHD data may link to the 2010 Census but be deleted from our analysis because that person's 2010 Census race or ethnicity data were edited or allocated. About 10% of the eligible workers in LEHD do not link to the 2010 Census, and an additional 10% did not have respondent-provided race and ethnicity data. We made no adjustments for these deletions in our statistical analyses.

[14] We make no use of employer identifiers. Thus, to keep the notation simple, *j* is an arbitrary index that ranges over the available earnings records for individual *i* in year *t*.

[15] Some Hispanic places of birth failed the minimum population criterion (1,000) for inclusion in our tables: Guinea-Bissau, Mayotte, and New Caledonia. These countries are included in "all other foreign born."

Instead, we coded all place of birth countries from Morocco on the west along the Mediterranean Sea to Syria on the north, then eastward to Iran and southward to Yemen.[16]

To better understand how the detailed race and ethnicity information contained in the 2010 Census checkbox and write-in responses relates to place of birth, we created a variable that records whether the country of birth from the Census Numident matches the country reported in the checkbox or write-in response from the 2010 Census. The country of birth on the Census Numident comes from the application for a Social Security Number, whereas the 2010 Census origin countries are self-identified. When the reported country in the checkbox or write-in responses on the 2010 Census matches the place of birth on the Census Numident, we code the individual as "Census detail/SSA POB Match."[17] We interact this variable with our 16 ethnicity by race categories to create a 32-category Hispanic ethnicity by race by Census detail/SSA POB match analysis variable, which we use to stratify both analysis samples. We do not report results for all 32 categories. Instead, we focus on the top X reported responses, where X ≤ 10, for each Census Numident place of birth. Fewer than 10 categories are listed when the sample for that category is too small to permit meaningful summaries.[18] In the tables we label rows with a 3-character tag. The first position is H for Hispanic ethnicity, N for Not Hispanic; the second position is the numbers 1, …, 8 for White, Black, American Indian/Alaska Native, Asian, Native Hawaiian/Pacific Islander, Some Other Race, 2+ Races (not SOR), and 2+ Races (SOR) in that order; and the third position is P for place-of-birth match, N for no place-of-birth match. We also label panels in the tables with a mnemonic that allows us to refer to the group in that panel (*e.g.*, FBH for Foreign-Born Hispanic) along with the sub-groups in that panel (*e.g.*, H1P for Hispanic, White, POB match). In the text, we cross-reference numerical data from the tables using these panel and row labels.

---

[16] This definition excludes Turkey, which would also be excluded if we applied the same rule used for Hispanic-origin countries using the results in the 2015 National Content Test (Matthews *et al.* 2017, Table 22) because less than 50% of persons who self-identified as Turkish also self-identified as MENA when given the opportunity to do so. On the other hand, our definition includes Israel, even though less than 50% of self-identified Israelis also self-identified as MENA on the 2015 National Content Test. There is no clean way to define MENA-origin countries comparable to our method of defining Hispanic-origin countries unless place of birth and self-identification data are available for the same person. Some MENA countries failed the minimum population criterion (1,000) for inclusion in our tabulation: Brunei, Oman, Qatar, United Arab Emirates, Gaza and West Bank. These birth countries are included in "all other foreign born."

[17] The 2010 Census detailed race and ethnicity variables have a corresponding country code for every Hispanic country, but some MENA source countries either did not have a response or were not coded. The following MENA country codes are not available on the 2010 Census: Algeria, Bahrain, Jordan, Kuwait, Libya, Saudi Arabia, Tunisia, and Yemen.

[18] Rows in the results tables are ordered from largest to smallest of these 32 categories based on the sum of the number of females and males in the row for that POB panel. Starting from the largest row, rows are included if the cumulative proportion of the within POB total is ≤ 0.99 and the sample size of the row is ≥ 100. Any remaining rows are collated into a single "All Other Responses" category. No POB panel may have more than 10 rows.

In addition to studying long-term-average real annual earnings, we also study regression-adjusted earnings, using personal characteristics (education, age, and years living in the U.S.), labor force attachment (number of years with any earnings), and hours of work. Previous work indicates that these are important predictors of long-term average real annual earnings. The regression adjustment models are run separately for each POB country.

Although information on hours of work and education are not available for the entire population, we assume the data are missing at random in the sense of Little and Rubin (2002) and impute the missing values conditional on all observed data. We imputed missing hours using information from the small subset of states (Minnesota, Oregon, Rhode Island, and Washington) for which hours data are reported. We estimated a least-squares regression model of log annual work hours at a given job as a function of a log earnings quartic, age quartic, race indicators, a foreign-born indicator, and NAICS 2017 industry sector indictors. If the worker has multiple jobs during the year, then the sum of earnings at all other jobs is included as an additional covariate. We estimated the imputation regression model separately for workers with different quarterly work patterns, dominant jobs, coincident jobs, and sex.[19]

Education is observed for respondents in the 2000 Census and all available years of the American Community Survey; approximately 20% of individuals have reported education. We imputed missing education using information about a person's observed characteristics (sex, place of birth, age, race, and ethnicity) as well as the characteristics of a person's job history such as the average earnings, modal industry, and characteristics of a person's co-workers and co-residents. The characteristics are used to form homogeneous cells of a minimum size within which the distribution of observed education values is used to impute missing education values.[20]

It is important to keep in mind that our methods are designed to isolate differences in adjusted long-term earnings *within recorded place of birth* between our 32-category classification of self-identified ethnicity, race, and POB match. That is, we are looking for differences in the measured outcomes that are associated with differences in the way individuals self-identify not differences related to their place of birth. We focus on these differences related to self-identification because they are at the heart of understanding meaningful changes in the way statistics about race and ethnicity may be reported and studied when the statistical standards for those classifications change under the 2024 revision of SPD 15.

---

[19] Hahn, Hyatt, and Janicki (2021) used a similar hours of work imputation methodology. That paper contains a detailed evaluation of the imputation and a comparison to hours statistics found in other data sources.
[20] McKinney *et al.* (2020) show that the conditional missing at random assumption, given the observed data, holds for education and that this method of imputation is reliable. Margins of error are not meaningfully inflated for sample sizes comparable to those in this paper.

Table 1 presents summary statistics for both samples and all variables used in our analyses. The statistics are reported for the four main classifications we use: Foreign-Born Hispanic, Born in the Middle East and North Africa, Other Foreign Born, and Born in the United States of America (excluding Puerto Rico).[21] In addition, for Sample 2, the table reports Foreign Born Parents From More Than One Group. For Sample 2, the categories refer to the place of birth of the biological parent with whom the person in Sample 2 was living when the 2000 Census was collected. All Sample 2 individuals were born in the U.S.

Because the individuals in Sample 1 are much older than those in Sample 2 and because long-term average annual earnings in Table 1 are not adjusted for differences in observable characteristics, comparisons across the samples must be done with care. Our measure of average annual earnings $w_i$ does not control for labor supply during the six-year eligible worker period, although we recognize that labor supply likely differs across groups. To address this, we incorporate years and hours worked measures into our analysis as right-hand side covariates. By controlling for labor supply as a covariate and not directly, we are able to estimate median average annual earnings for each group, either accounting for or not accounting for labor supply differences.

As expected, mean hours paid during the year and the average number of years with positive reported earnings are higher for males than for females. For example, female workers born in Hispanic countries are paid on average for 1,364 hours during an active year and are inactive on average over the six-year period for 1.09 years, compared with 1,676 hours and 0.86 years for foreign born Hispanic males. Hispanic-born women work about one full-time week per year less than native-born female workers (1,364 compared with 1,404 hours paid per year) when active and are also more likely to spend a full year inactive (1.09 compared with 0.81 inactive years). Overall, workers born in MENA countries have somewhat less attachment to the labor force than workers born in Hispanic countries, although the differences are not large.

In addition to the labor supply differences across groups, some education patterns are worth mentioning. A very large percentage of those born in Hispanic countries have less than a high-school education (44% of females and 49% of males) as compared with the much lower percentages for those born in the U.S. (8% of females and 11% of males) even though the samples have essentially the same average age (about 40 years old in 2010). A greater percentage of the USA-born children of Hispanic-born parents also have less than a high-school education (23% for females and 24% for males) as compared to the children of USA-born parents (13% for females and 15% for males) even though both samples are essentially the same age (26 years old). The narrowing of the gap between the 1st-generation immigrants (parents) and 2nd generation (children of foreign-born parents) is a common feature of intergenerational data.

[21] We call out the exclusion of Puerto Rico explicitly because the LEHD and 2010 Census data include records for persons living in Puerto Rico, but we deleted them from our universe. Note that individuals born in Puerto Rico and living in the U.S. are included in the Foreign-born Hispanic category.

A greater percentage of MENA-born persons in Sample 1 have at least a bachelor's degree (48% of females and 51% of males) as compared to the percentages for USA-born persons (30% for females and 27% for males). This larger differential in higher education persists for the children of MENA-born parents (40% for females and 34% for males) as compared to the children of USA-born parents (26% for females and 22% for males).

## 4. Results

The results in Table 2 show how adjusted long-term average earnings change for Sample 1 workers as we sequentially add regression controls for person characteristics (PC), workplace characteristics (PCW) and hours worked (PCWH). They also provide a comparison group for Sample 2 workers. Tables 2, 3, and 4 follow a similar layout. The rows are organized into groups based on the respondent's own (Tables 2 and 3) or their parent's (Table 4) place of birth. We reference statistics in these tables using the Row Label column.

In Table 2 we show four different, mutually exclusive groups based on the respondent's place of birth: Foreign-Born Hispanic, Middle East and North Africa, Other Foreign Born, and USA-Born. The first row within each POB group contains the largest population (females plus males) ethnicity by race by Census Detail/SSA POB match sub-group. The remaining rows are in decreasing size order, where the size is determined by the proportion of the total panel population in that row. For the earnings results, the largest population, or first row, in each POB group is the reference sub-group. The Mean Log Earnings column shows average log long-term real earnings for workers in the largest sub-group. For every other row the table displays the difference relative to the largest sub-group. For example, female foreign-born Hispanic average log long-term real earnings is 9.360 for the Hispanic, White, Census Detail/SSA POB Match (FBH H1P) reference sub-group, and the next largest sub-group Hispanic, Some Other Race, Census Detail/SSA POB Match (FBH H6P) has long-term average earnings -0.046 log points relative to the reference sub-group. The results for males are presented in a similar format using the same reference sub-group as females.

For both females and males, we estimate separate earnings regressions within each POB group, controlling for three sets of characteristics, with each regression model building on the previous model as you move to the right in the table. In Table 2 only, we present the PC and PCW models to better understand the relative impact of the different sets of characteristics. The PC model includes controls for education and quadratic functions of age and years in the U.S. in 2010. The PCW model adds controls for the dominant industry and region and the number of full and partial years worked. The PCWH model adds a quadratic function of average annual hours worked.

In our first adjusted model, PC, we control only for person characteristics. Including person characteristics generally reduces the difference in earnings relative to the reference category although the impacts are uneven and vary by POB group, ethnicity, and race category. For example, the unadjusted difference in mean log earnings for Not Hispanic, White, females in the Other Foreign-Born group relative to the reference sub-group is -0.085 log points (OFB N1N), while the difference controlling for person characteristics (PC) is -0.095 log points (OFB N1N). Including controls for the type of workplace and the number of years worked (PCW) reduces the gap from -0.095 in the PC model to -0.044 log points in the PCW model (OFB N1N). Including controls for hours worked reduces the gap further from -0.044 log points in the PCW model to 0.029 log points in the PCWH model (OFB N1N).

When interpreting the results, the PC and PCW model estimates typically move in predictable ways, however in many cases when controlling for hours of work (PCWH), the model estimates may be counterintuitive. For example, consider the estimates for the Foreign-Born Hispanic females in the Hispanic, Black, Census Detail/SSA POB Match (FBH H2P) sub-group, the earnings gap actually increases from -0.039 log points to -0.138 log points when we control for hours worked. This result implies that for this ethnicity, race, POB-match sub-group, individuals work substantially more hours than the reference sub-group, and once we control for this difference in the PCWH model, the earnings gap increases. Although the PCWH model is our preferred specification there are notable differences between the PCW and the PCWH model results. A case can also be made for using the PCW model results, especially if both earnings differences and hours worked are driven by similar underlying processes, for example, discriminatory practices in the labor market.[22]

---

[22] In these analyses we consider median real average annual earnings, the logarithm of real average annual earnings, mean hours paid, mean full years worked, mean partial years worked, and mean inactive years. In earlier work (McKinney, Abowd and Janicki, 2022) we performed quantile regression to assess the distributional differences in the relationship of covariates to long-term annual earnings. That work distinguished native and foreign-born workers and used the same race and ethnicity categories considered in this paper. We found substantial differences in the effect of labor-force attachment in the quantiles below the median. Our conclusion was "[a]t the bottom of the earnings distribution, differences in earnings across groups are largely due to observable characteristics suggesting that workers at the bottom of the earnings distribution may have the clearest path to improving their position. Increasing hours paid, changing employers, and attaining additional education, while difficult in many cases, are some of the standard pathways to higher real earnings. For workers above the median, differences in the return to characteristics is the dominant component. The pathway to reducing differences in the returns to observable characteristics across demographic groups is less clear. (p. 1924)." See also Tables 2A and 2B therein. Because the differences in labor force attachment (measured as either average annual hours or years of zero employment) are substantial in the lower tail of the distribution by race and ethnicity but not at the median, we don't further analyze those effects in this paper.

Table 3 reports the detailed results for long-term earnings differentials for Sample 1 individuals with and without adjustment by sex for each place of birth country with at least 1,000 persons. We discuss the results for the largest immigrant birth countries in the Hispanic and MENA categories.

Mexico is the largest Hispanic immigrant source country in our data by a substantial margin. Essentially all immigrants in Sample 1 born in Mexico self-identify as Hispanic (MEX 99.3%).[23] Similarly, nearly all Mexican-born immigrants (MEX, 96.6%), self-declared Mexico as a checkbox or write-in response on the 2010 Census. Our reference sub-group (MEX H1P, 50.0%) self-identified as Hispanic and White in addition to indicating Mexico on either the race or ethnicity question in the 2010 Census. Compared to this reference sub-group, the sub-group that self-identified as Hispanic and Some Other Race (MEX H6P, 42.2%) and the sub-group that self-identified as 2+ Races (SOR), *i.e.*, including Some Other Race (MEX H8P, 2.8%), have slightly lower adjusted log earnings for both females and males. Notice also that the sub-group that self-identifies as Hispanic and White but does not check or write-in Mexico on the 2010 Census (MEX H1N, 1.3%) is indistinguishable from the reference sub-group (less than 0.01 log point difference in adjusted log earnings for females and males). Similarly, the sub-group that self-identifies as Hispanic and Some Other Race but does not select or write-in Mexico on the 2010 Census (MEX H6N, 1.3%) is indistinguishable from those that self-identified as Hispanic, Some Other Race, and did check or write-in Mexico (again, less than -0.01 log point difference in adjusted earnings).

Prior research strongly suggests that all five of the top sub-groups (MEX H1P-H1N)—97.5% of all Mexican immigrants—will self-identify as Hispanic or Hispanic and White, and probably select the Mexican checkbox on the proposed single-question SPD 15 (Matthews *et al*. 2017, Table 5). About half of persons who self-identified as Hispanic, can be expected to self-identify as Hispanic alone in the single question format. Only a negligible percentage can be expected to self-identify as Hispanic and Some Other Race. If so, then the race and ethnicity information on Mexican immigrants from the single-question format will lose some granularity—the difference between those who self-identify Hispanic and White versus Hispanic and Some Other Race in the two-question format.[24] Our results suggest the magnitude of this heterogeneity is not large for Mexican-born immigrants. For Mexican-born females who self-identify as Hispanic, Some Other Race, the adjusted difference is -0.033 log points (MEX H6N) compared to the unadjusted difference of -0.583 log points between Hispanic, White, Mexican-born females and Not Hispanic, White USA-born females (Table 3 MEX H1P – Table 2 USA N1N). For Mexican-born males who self-identify as Hispanic, Some Other Race, the adjusted difference is -0.041 log points (MEX H6N) compared to the unadjusted difference of -0.360 log points between Hispanic, White, Mexican-born and Not Hispanic, White USA-born males (Table 3 MEX H1P – Table 2 USA N1N).

[23] In reporting results for Tables 3 and 4 we use the panel label (in this case MEX) and the row label (for the reference row in the Mexican panel H1P) to help readers find the results we are discussing.

[24] We remind the reader that there is no social-political construct behind the category Some Other Race. We call out the distinction here because the choice "Hispanic, Some Other Race" is allowed under both the 1997 and 2024 revisions of SPD 15, but the 1997 requirement to answer a separate question about Hispanic ethnicity has been eliminated in the 2024 revision. Most published historical tables embody the coding rules discussed in the introduction. While we did not use those coding rules, we did tabulate results for individuals who complied with the 1997 format and separately answered Some Other Race. Matthews *et al.* 2017 shows that virtually no one who selects Hispanic in the single-question format will also select Some Other Race. Hence, a recoding of historical data might eliminate the Some Other Race response for these individuals.

The second most populous group of foreign-born Hispanics in Sample 1 is those born in Puerto Rico. Before discussing these results, we remind the reader that all persons living in the Commonwealth of Puerto Rico are eligible for U.S. Social Security Numbers; however, only those living in the U.S. are included in our universe. (We did not use the recently added Puerto Rican data in the LEHD infrastructure files.)

Essentially all Puerto Rican-born U.S. residents self-identify as Hispanic (PR, 96.8%). The largest sub-group, hence, our reference sub-group, self-identify as Hispanic, White, and checked or wrote-in Puerto Rico in the race or ethnicity responses on the 2010 Census (PR H1P, 59.7%). In terms of adjusted log earnings differentials, there is more heterogeneity among the Puerto Rican-born immigrants in our Sample 1. First, note that a substantial percentage self-identify as Hispanic, Black, and check or write-in Puerto Rico (PR H2P, 6.5%). This group has substantially lower adjusted log earnings (PR H2P, -0.134 log points for females and -0.131 log points for males) than the reference sub-group (PR H1P). Second, both the second largest sub-group—those who self-identify as Hispanic, Some Other Race, and checked or wrote-in Puerto Rico (PR H6P, 23.8%)—and the fourth largest sub-group—those who self-identified as Hispanic, 2+ Races including Some Other Race, and checked or wrote-in Puerto Rico (PR H8P, 2.5%)—and the seventh largest sub-group—those who self-identify as Hispanic, Some Other Race, but did not check or write-in Puerto Rico (PR H6N, 1.3%)—have at least -0.049 log point lower adjusted long-term earnings for females and males than the reference sub-group (PR H1P). Finally, we note that Puerto Rican-born immigrants who self-identified as Hispanic and White but who did not check or write-in Puerto Rico (PR H1N, 1.8%) have slightly larger adjusted log earnings than the reference group (0.027 log points for females and 0.023 log points for males compared to PR H1P).

Because Puerto Ricans who self-identify as Hispanic and Black have substantially different long-term real earnings than Puerto Ricans who self-identify as Hispanic and White, Some Other Race or 2+ Races (SOR), the likely consequences of using the single-question format are more difficult to predict. If most of the Puerto Rican-born persons who self-identified as Hispanic and Black in the two-question format continue to do so in the single-question format, then future analysts will be able to study their distinct labor market outcomes. But if most elect to self-identify as Hispanic alone, then decompositions like the ones we do in this paper will no longer be possible. Previous research has shown that approximately the same percentage of those who self-identify as Hispanic will also self-identify as Hispanic and Black in either the single-question or two-question format (Matthews *et al*. 2017, Table 7). Essentially no Hispanics self-identify as Hispanic and Some Other Race in the single-question format (*ibid*., Table 7).

The third most populous group of foreign-born Hispanics in Sample 1 were born in El Salvador (ELS). Within this group the largest sub-group and, hence, our reference sub-group is persons who self-identify as Hispanic, Some Other Race, and either checked or wrote-in El Salvador on the 2010 Census (ELS H6P, 46.5%). Only two sub-groups within El Salvador have markedly different adjusted long-term earnings differentials. As we saw for Puerto Rican-born persons, El Salvadorian immigrants who self-identify as Hispanic, Black, and checked or wrote-in El Salvador (ELS H2P) had lower adjusted log earnings than the reference group (-0.084 log points for females and -0.082 log points for males). Those who self-identify as Not Hispanic, White, and did not check or write-in El Salvador (ELS N1N) have adjusted differentials of 0.170 log points for both females and males, indicating that they closely resemble USA-born persons who identify as Not Hispanic and White.

The expected consequences of switching from the two-question to the single-question format for El Salvadorian immigrants are similar to the consequences for Puerto Ricans. We expect those who self-identified as Not Hispanic and White would continue to do so. The uncertainty surrounds those who would self-identify as Hispanic and Black or White vs. Hispanic alone.

The fourth most populous group of foreign-born Hispanics in Sample 1 were born in Cuba (CUB). Within this group the largest sub-group and, hence, our reference sub-group self-identified as Hispanic, White, and checked or wrote-in Cuba (CUB H1P, 87.7%). The heterogeneity of adjusted long-term earnings is much stronger within the Cuban immigrants. Those who self-identified as Hispanic, Black, and checked or wrote-in Cuba have lower adjusted earnings (CUB H2P, -0.162 log points for females and -0.144 log points for males). Those who self-identified as Not Hispanic, White, and did not check or write-in Cuba had higher adjusted earnings but not as high as in most other immigrant groups (CUB N1N, 0.097 log points for females and 0.092 log points for males).

The expected consequences of switching to the single-question format are again similar to Puerto Rico, but potentially more pronounced. More heterogeneity is masked for individuals who self-identify as Hispanic alone.

Immigrants from all countries classified as Hispanic in this paper who self-identified as Not Hispanic and White rarely listed their birth country, as recorded on their Social Security Number applications, as one of the detailed checkbox or write-in responses on the 2010 Census. Immigrants from the Hispanic source countries used in this paper who self-identified as Not Hispanic, White, and checked or wrote-in the same country as their SSA place of birth are therefore in the "All Other Responses" category of our tables because of this paucity of data. We do show data for immigrants from Hispanic countries who self-identified as Not Hispanic and White but did not include their birth country in either the checkboxes or write-ins on the 2010 Census for the larger origin-country populations. Except for those born in Spain, immigrants born in the countries we labeled Hispanic who self-identified as Not Hispanic and White had unadjusted and adjusted long-term earnings differences compared to those who self-identified as Hispanic and White or Hispanic and Some Other Race that were similar to the those of USA-born persons who self-identified as Not Hispanic and White compared to USA-born persons who self-identified as Hispanic and White. With respect to characterizing labor market outcomes for immigrants, therefore, there is some information in the self-declaration Not Hispanic and White for those born in countries we classified as Hispanic. Although the data are sparser, a similar pattern holds for those born in Hispanic countries who self-identify as Not Hispanic and Asian. The pattern is less clear for those born in these countries who self-identify as Not Hispanic and Black. Our results suggest that these differences would still be estimable under the single-question format.

Summarizing the results for MENA immigrants is not as complicated by comparisons of the single-question and two-questions formats because in the current two-question format self-identifying as MENA requires using a checkbox or write-in within one of the five official race categories or Some Other Race. That said, the same estimated percentage of persons self-identify as MENA alone in either format (Matthews *et al.* 2017, Table 5). Only limited information was published about the combinations of MENA and other race groups in the results of the 2015 National Content Test that we have been citing. Very few MENA immigrants self-identify as Hispanic. The largest proportion of MENA-born persons self-identifying as Hispanic immigrated from Algeria (ALG, 6.7%) and the smallest from Iran (IRN, 0.2%). Only in Algeria, Israel, Morocco, and Saudi Arabia did any sub-groups including self-identified Hispanics meet the population threshold for inclusion in our tables. Consequently, the discussion of results for MENA-born immigrants focuses exclusively on those who self-identified as Not Hispanic.

The largest MENA-born immigrant group in Sample 1 was born in Iran. Among Iranian immigrants, the largest sub-group and, hence, our reference sub-group, self-identifies as Not Hispanic, Some Other Race, and checked or wrote-in Iran on the 2010 Census (IRN N6P, 38.0%). The second largest sub-group self-identifies as Not Hispanic, White, and did not check or write-in Iran (IRN N1N, 31.9%). The differences in adjusted long-term earnings between these first two sub-groups are small (-0.014 for females and -0.001 for males). The next largest sub-group self-identified as Not Hispanic, 2+ Races (SOR), and either checked or wrote-in Iran (IRN N8P, 13.6%). The differences in adjusted long-term earnings relative to the reference sub-group may be larger for this sub-group (-0.030 log points for females and -0.044 log points for males). The fourth largest sub-group self-identifies as Not Hispanic, Some Other Race, and did not check or write-in Iran (IRN N6N, 13.6%). For this sub-group the differences in adjusted long-term earnings are pronounced (-0.135 log points for females and -0.163 for males). Similarly, the next largest sub-group, which identifies as Not Hispanic, 2+ Races (SOR), and did not check or write-in Iran (IRN N8N 1.5%), also has large differences in adjusted long term earnings (-0.184 for females and -0.168 for males).

Since MENA self-identifications can only occur via checkboxes and write-ins in the current two-question format, the striking difference in sub-groups noted for Iranian immigrants are likely to present some challenges in the proposed single-question format for the style of comparisons used in this paper. When there is no explicit MENA category (as in the current two-question format), 80.7% of Iranians self-identify as White, 13.8% as Some Other Race, with the balance in one of the other current categories (Black, American Indian/Alaska Native, Asian, or Native Hawaiian/Pacific Islander) (Matthews *et al*. 2017, Table 22). Conversely, when there is a MENA category (as in the proposed single-question format), only 14.7% of Iranians self-identify as White, while 85.0% self-identify as MENA, and only 2.1% self-identify as Some Other Race (*ibid*. Table 22). The marked differences in adjusted long-term earnings that we find between Iranian immigrants who self-identified as White or Some Other Race *and indicated Iran in the checkboxes or write-ins*, compared to those who self-identified as Some Other Race (either alone or in combination) *and did not indicate Iran in the checkboxes or write-ins* will probably not be detectable in the single-question format without ancillary information on place of birth.

The second largest MENA-born group in Sample 1 was born in Israel. Among Israeli immigrants, the largest sub-group and, hence, our reference sub-group, self-identified as Not Hispanic, White, and did not check or write-in Israel (ISR N1N, 79.6%). Very few Israeli-born immigrants checked or wrote-in Israel (only ISR N6P, 4.4%, and ISR N8P, 0.5%). The second largest sub-group self-identified as Not Hispanic, Some Other Race, and did not check or write-in Israel (SR N6N 11.5%). There are large differences in adjusted long-term earnings between this sub-group and the reference sub-group (-0.070 log points for females and -0.179 log points for females). We find even larger differences in adjusted long-term earnings for the other sub-groups of Israeli immigrants that self-identified as Not Hispanic and Some Other Race in any form.

The large differences in adjusted long-term earnings between Israelis who self-identify as White and those who self-identify as Some Other Race may also be difficult to detect in the proposed single-question format. When self-identified Israelis responded to the race question with no MENA category, 90.9% self-identified as White and 4.8% as Some Other Race. When there was a MENA category, only 49.6% self-identified as White, while 38.9% self-identified as MENA, and 10.6% self-identified as Some Other Race. (Matthews *et al*. 2017, Table 22).

The third largest MENA-born immigrant group in Sample 1 is Egyptian. Among Egyptian immigrants the largest sub-group and, hence, our reference sub-group, self-identified as Not Hispanic, Some Other Race, and checked or wrote-in Egypt (EGY N6P, 43.2%). The second largest sub-group self-identified as Not Hispanic, White and did not check or write-in Egypt (EGY N1N, 36.9%). For this sub-group adjusted long-term earnings differentials were meaningful but not large (0.051 log points for females and 0.080 log points for males). Interestingly, the third largest sub-group, which self-identified as Not Hispanic, Some Other Race, and did not check or write-in Egypt (EGY N6N, 13.7%), had adjusted long-term earnings differentials that were very similar to the White sub-group (0.046 log points for females and 0.058 log points for males). In contrast, the fourth largest sub-group, which self-identified as Not Hispanic, Black, and did not check or write-in Egypt (EGY N2N, 2.0%), had large, negative long-term earnings differentials (-0.111 log points for females and -0.112 log points for males).

When answering in the two-question format, 91.0% of self-identified Egyptians chose the White category, 2.3% chose Black, and 2.3% chose Some Other Race (Matthews *et al*. 2017, Table 22). In the single-question format, 91.5% chose MENA, 8.4% chose White, and 2.3% chose Some Other Race (*ibid.*). If these choices carry over to the proposed single-question format, the adjusted earnings differentials between those who self-identify as White or Some Other Race and those who self-identify as Black will no longer be estimable even if immigrant country of origin information is available. Since more than half of Egyptian immigrants did not self-identify as Egyptian in their census responses, this may not be a serious limitation unless the presence of a MENA category in the single-question format results in a greater proportion of Egyptian immigrants choosing MENA alone.

The fourth largest MENA-born immigrant group is Lebanese. Among Lebanese immigrants, the largest sub-group and, hence, our reference category, self-identified as Not Hispanic, White, and did not check or write-in Lebanese (LEB N1N, 58.3%). The second largest sub-group self-identified as Not Hispanic, Some Other Race, and did not check or write-in Lebanese (LEB N6N, 22.6%). Long-term earnings differentials in the second sub-group are substantial for males (-0.041 log points for females and -0.102 log points for males). The third largest sub-group self-identified as Not Hispanic, Some Other Race, but did check or write-in Lebanese (LEB N6P, 11.7%). Long-term earnings differentials for this third sub-group, which only differs from the second sub-group by self-identifying as Lebanese, were substantial (-0.072 log points for females and -0.123 log points for males). The fourth-largest sub-group self-identified as Not Hispanic, 2+ Races (SOR) (LEB N8P, 3.2%). This sub-group is indistinguishable from the third sub-group in the magnitude of long-term earnings differentials (-0.073 log points for females and -0.115 log point for male). Finally, we note that Lebanese immigrants who self-identify as Not Hispanic, Black, and did not check or write in Lebanese are a small portion of the total (LEB N2N, 0.5%) but have enormous negative adjusted long-term earnings differentials (-0.376 log points for females and -0.651 log points for males).

When answering in the two-question format, 91.7% of self-identified Lebanese chose White, while only 1.9% chose Some Other Race, and too few chose Black to measure reliably (Matthews *et al*. 2017, Table 22). In the single-question format, 34.0% chose White, and 69.0% chose MENA (*ibid*.). Like the Egyptian-born immigrants, most Lebanese-born immigrants do not self-identify as Lebanese in the current two-question format. Hence, it is difficult to predict the consequences of the proposed single-question format on these immigrants.

We now turn to Table 4. Given the results discussed above for 1st generations immigrants (workers born abroad), another population of interest is 2nd generation immigrants (workers born in the USA, with at least one parent born abroad). For 2nd generation workers we would like to understand how similar their responses and earnings across ethnicity-race-SSA POB Match categories are to 1st generation immigrants from the same POB group. Another useful comparison group we also use is to the same ethnicity-race 3rd generation and above native-born workers.

Identifying 2nd generation workers is more complicated than identifying 1st generation workers and requires additional information on the place of birth of the biological parents. We use responses from the parents on the 2000 Decennial Census to identify the biological parents when the sample member was 15-18 years old. At that point in time either one, two or no biological parents may be present in the home. We retain in Sample 2 workers with at least one biological parent in the home. For a worker to be in the Hispanic group; the person must have had one Hispanic-born parent in the home; two parents in the home, one of whom is from a Hispanic source country; or two parents in the home, both of whom are from a Hispanic source country. The MENA and Other Foreign-Born groups are processed similarly. We have a final category that captures the remaining possibilities, where for example one parent is from the Hispanic group and another parent is from the MENA group.

Family structure at the age of the respondent in 2000 (15-18) matters for the responses received on the ethnicity and race questions. For example, workers with one foreign-born parent and one USA-born parent are noticeably more likely to report Not Hispanic, White (2P1FBH N1N), the largest category for 3rd generation and above workers. For MENA workers with one foreign-born and one USA-born parent, the percentage of workers who report in the largest category Not Hispanic, White is 80.3% in Table 4 (2P1MENA N1N), while for workers with two MENA parents the equivalent percentage is 44.8% (2MENA N1N). We also find a relatively large difference in the proportion of self-identified Not Hispanic, White between the Other

Foreign Born parent groups (2P1OFB N1N 78.1% vs. 2OFB N1N 21.1%) and between the Hispanic country parent groups (2P1FBH N1N 17.5% vs 2FBH N1N 1.7%). For the workers with only one foreign-born parent in the home the proportion Not Hispanic, White is typically larger than for workers in the same POB group with two foreign born parents, but less than those with two parents, one of whom was born in the U.S.

Long-term average earnings also differ noticeably by family structure both within and across POB groups. For males, log earnings for the largest sub-group Hispanic, White, Census Detail/SSA POB Match within the Hispanic origin parent groups is highest for workers with one USA-born parent (2P1FBH H1P 10.01), but this is the exception. For all other groups (females with Hispanic-origin parents, and both sexes for MENA and Other Foreign Born parent groups), workers in the largest ethnicity-race-POB Match category who have two foreign-born parents have the highest long-term average earnings. This result is similar to the earnings pattern present in most studies of immigrant assimilation where 2$^{nd}$ generation workers outperform the 3$^{rd}$ generation and above.

We can also compare long-term average earnings for Hispanic 2$^{nd}$ generation male workers with two foreign-born parents to 3$^{rd}$ generation and above long-term average earning outcomes. For Hispanic White, Census Detail/SSA POB Match male workers with two foreign-born parents, log long-term average earnings is 9.995 (2FBH H1P), while for 3$^{rd}$ generation Hispanic, White male workers with two parents long-term log average annual earnings is (10.18 - 0.201 = 9.979, 2US N1N – 2US H1N). Similar patterns can be found for other POB groups in Table 4.

We focus now on comparing the results in Table 2 for 1$^{st}$ generation workers with 2$^{nd}$ generation workers with two foreign-born parents in Table 4, keeping in mind that the workers in Table 2 are significantly older on average (approximately 40 years old in 2010) than the workers in Table 4 (approximately 26 years old in 2010). We begin with the Hispanic POB group. Although the sample composition differs, ethnicity and race reporting is about the same between the older 1$^{st}$ generation workers and the younger 2$^{nd}$ generation workers with the same top two categories and a similar cumulative proportion for the top two categories of 0.863 (Table 2 FBH H6P) for 1$^{st}$ generation immigrants and 0.880 (Table 4 2FBH H6P) for 2$^{nd}$ generation immigrants. The differences are somewhat larger for the MENA group although the top two categories are the same, the cumulative proportion is 0.678 (Table 2 MENA N6N) for the 1$^{st}$ generation and 0.752 (Table 4 2MENA N6N) for the 2$^{nd}$ generation with most of the difference due to a higher share of 2$^{nd}$ generation workers reporting Not Hispanic, Some Other Race (Table 4 2MENA N6N 0.304 vs Table 2 MENA N6N 0.217). For the Other Foreign-Born group, the top two ethnicity and race categories and the cumulative proportion for the top two categories is roughly similar— 0.773 (Table 2 OFB N1N) for the 1$^{st}$ generation and 0.823 (Table 4 2OFB N1N) for the 2$^{nd}$ generation—although the share of workers in each category is noticeably different. The share of 2nd generation Not Hispanic, Asian workers is 0.611 (Table 4 2OFB N4N) versus 0.477 (Table 2 OFB N4N) for 1st generation workers. It isn't entirely clear what is driving this result except perhaps changing trends in immigrant composition and the older average age for workers in Sample 1.

Long-term average earnings for 2nd generation workers show relatively large variability across ethnicity and race categories within POB groups, with variability at least as large and often substantially larger than for 1st generation workers. For female workers in the Hispanic POB groups in Tables 2 and 4, the second largest reported ethnicity and race category is Hispanic, Some Other Race, Census Detail/SSA POB Match (H6P). The difference in unadjusted long-term average earnings relative to the reference group Hispanic, White, Census Detail/SSA POB Match (H1P) is -0.046 (Table 2 FBH H6P) for 1st generation females and -0.093 (Table 4 2FBH H6P) for 2nd generation females with two foreign-born Hispanic parents. It is even more important to know whether a person self-identifies White or Some Other Race for 2nd generation female workers than it is for 1st generation female workers. For Hispanic POB males, the 2nd generation differences are similar across the top three ethnicity-race categories, but the unadjusted difference is very large for the Hispanic, White category, -0.067 (Table 2 FBH H1N) for 1st generation males and -0.224 (Table 4 2FBH H1N) for 2nd generation males.

The pattern for Hispanic POB workers repeats itself for the MENA and the Other Foreign Born groups. Female 2nd generation workers tend to have larger earnings dispersion across ethnicity and race sub-groups within POB group than 1st generation female workers, while the converse is true for male workers. This implies that the loss of information in the single-question format for 2nd generation male workers may be less than for 2nd generation females when, for example, a worker that would have self-identified in two race categories in the two-question format does not do so using the single-question format.

## 5. Conclusion

Changing the way a society measures its racial and ethnic diversity happens regularly. New standards for the collection and reporting of self-identified demographic data can produce statistical summaries that better reflect the current population of the United States. But these changes can also make historical comparisons more difficult. Our paper demonstrates some of these challenges. In measuring the long-term earnings of Hispanics, the proposed single-question format may complicate comparisons to analyses from the current two-question format because individuals must self-identify into White, Black, or other categories in addition to Hispanic. For individuals who volunteered self-identification as Middle Eastern or North African in the two-question format, the statistics based on the proposed single-question format may simply not be comparable to any previous analyses. Providing country of birth data for those born outside the United States may permit more comparable historical studies for immigrants.

# Tables

**Table 1 - Summary Statistics for Samples 1 and 2**

| | Sample 1 | | Sample 2 | |
|---|---|---|---|---|
| | Females | Males | Females | Males |
| Total Sample Size | 44,640,000 | 43,320,000 | 3,614,000 | 3,582,000 |
| | Foreign-Born Hispanic | | | |
| Median Average Annual Earnings | 16,970 | 28,640 | 22,960 | 27,360 |
| Mean Log Average Annual Earnings | 9.35 | 9.93 | 9.69 | 9.90 |
| Mean Hours Paid | 1,364 | 1,676 | 1,438 | 1,593 |
| Mean Full Years Worked | 3.92 | 4.21 | 4.18 | 4.23 |
| Mean Partial Years Worked | 0.99 | 0.93 | 1.01 | 1.03 |
| Mean Inactive Years | 1.09 | 0.86 | 0.82 | 0.74 |
| Mean Age (2010) | 40.42 | 40.68 | 26.42 | 26.43 |
| Mean Years in U.S. (2010) | 18.96 | 19.81 | | |
| Less than High School | 0.44 | 0.49 | 0.23 | 0.24 |
| High School | 0.21 | 0.21 | 0.28 | 0.30 |
| Some College | 0.21 | 0.18 | 0.33 | 0.32 |
| Bachelor or Higher | 0.14 | 0.12 | 0.17 | 0.13 |
| Sample Size | 2,610,000 | 2,777,000 | 189,000 | 190,000 |
| | Born in the Middle East and North Africa | | | |
| Median Average Annual Earnings | 19,220 | 34,390 | 29,980 | 33,840 |
| Mean Log Average Annual Earnings | 9.54 | 10.21 | 9.97 | 10.11 |
| Mean Hours Paid | 1,312 | 1,594 | 1,426 | 1,547 |
| Mean Full Years Worked | 3.77 | 4.10 | 4.05 | 4.02 |
| Mean Partial Years Worked | 0.95 | 0.90 | 1.04 | 1.04 |
| Mean Inactive Years | 1.27 | 1.00 | 0.92 | 0.95 |
| Mean Age (2010) | 40.25 | 40.84 | 26.37 | 26.40 |
| Mean Years in U.S. (2010) | 17.62 | 17.95 | | |
| Less than High School | 0.11 | 0.11 | 0.09 | 0.11 |
| High School | 0.16 | 0.15 | 0.21 | 0.23 |
| Some College | 0.25 | 0.23 | 0.31 | 0.31 |
| Bachelor or Higher | 0.48 | 0.51 | 0.40 | 0.34 |
| Sample Size | 208,000 | 302,000 | 15,000 | 16,000 |
| | Other Foreign Born | | | |
| Median Average Annual Earnings | 28,250 | 43,930 | 30,260 | 33,370 |
| Mean Log Average Annual Earnings | 9.93 | 10.44 | 10.01 | 10.12 |
| Mean Hours Paid | 1,526 | 1,729 | 1,485 | 1,593 |
| Mean Full Years Worked | 4.29 | 4.48 | 4.32 | 4.31 |
| Mean Partial Years Worked | 0.79 | 0.74 | 0.93 | 0.95 |
| Mean Inactive Years | 0.92 | 0.78 | 0.75 | 0.74 |
| Mean Age (2010) | 40.22 | 40.25 | 26.43 | 26.45 |
| Mean Years in U.S. (2010) | 17.82 | 18.04 | | |
| Less than High School | 0.12 | 0.12 | 0.12 | 0.13 |
| High School | 0.18 | 0.17 | 0.24 | 0.26 |
| Some College | 0.27 | 0.24 | 0.31 | 0.31 |
| Bachelor or Higher | 0.43 | 0.47 | 0.33 | 0.29 |
| Sample Size | 4,256,000 | 4,015,000 | 256,000 | 256,000 |
| | Born in the United States of America (excluding Puerto Rico) | | | |
| Median Average Annual Earnings | 26,730 | 40,160 | 23,570 | 30,490 |
| Mean Log Average Annual Earnings | 9.81 | 10.26 | 9.70 | 10.00 |
| Mean Hours Paid | 1,404 | 1,626 | 1,367 | 1,558 |
| Mean Full Years Worked | 4.39 | 4.49 | 4.22 | 4.33 |
| Mean Partial Years Worked | 0.80 | 0.79 | 0.96 | 0.96 |
| Mean Inactive Years | 0.81 | 0.72 | 0.82 | 0.71 |
| Mean Age (2010) | 39.97 | 39.93 | 26.42 | 26.46 |
| Mean Years in U.S. (2010) | | | | |
| Less than High School | 0.08 | 0.11 | 0.13 | 0.15 |
| High School | 0.26 | 0.30 | 0.28 | 0.32 |
| Some College | 0.35 | 0.32 | 0.34 | 0.32 |
| Bachelor or Higher | 0.30 | 0.27 | 0.26 | 0.22 |
| Sample Size | 37,570,000 | 36,230,000 | 3,148,000 | 3,115,000 |
| | Foreign Born Parents From More Than One Group | | | |
| Median Average Annual Earnings | | | 30,090 | 31,230 |
| Mean Log Average Annual Earnings | | | 9.99 | 10.06 |
| Mean Hours Paid | | | 1,506 | 1,584 |
| Mean Full Years Worked | | | 4.26 | 4.15 |
| Mean Partial Years Worked | | | 0.96 | 1.02 |
| Mean Inactive Years | | | 0.78 | 0.83 |
| Mean Age (2010) | | | 26.40 | 26.41 |
| Mean Years in U.S. (2010) | | | | |
| Less than High School | | | 0.14 | 0.15 |
| High School | | | 0.24 | 0.25 |
| Some College | | | 0.31 | 0.31 |
| Bachelor or Higher | | | 0.31 | 0.27 |
| Sample Size | | | 5,900 | 5,900 |

Notes: Sample 1 consists of 87.97 million workers and Sample 2 consists of 7.196 million workers. Workers in both samples have at least one year with positive earnings during the six-year period from 2010 to 2015. Zero earning years are included in the earnings measures. Sample 1 includes both foreign and native-born workers age 25 to 55 (inclusive) in 2010, while

Sample 2 includes only native-born workers age 25 to 28 (inclusive). Place of birth is determined using the Census Numident. For Sample 2, the place of birth group is assigned using the parent's place of birth. The Hispanic place of birth group is the set of countries where at least 50% of foreign-born persons in Sample 1 report Hispanic on the 2010 Census. The full list of Hispanic and MENA countries can be found in Table 3.

## Table 2 – Mean Log Long-Term Average Annual Earnings by Sex, Place of Birth Group, Ethnicity, Race, and Census Detail/SSA Place of Birth Match

| Row Label | Ethnicity - Race - SSA Census Detail/SSA POB Match | Propor-tion | Cumula-tive Pro-portion | Females: Mean Log Earnings | Females: Adjusted Log Earn PC | Females: Adjusted Log Earn PCW | Females: Adjusted Log Earn PCWH | Males: Mean Log Earnings | Males: Adjusted Log Earn PC | Males: Adjusted Log Earn PCW | Males: Adjusted Log Earn PCWH |
|---|---|---|---|---|---|---|---|---|---|---|---|
| FBH | Foreign-Born Hispanic \| Sample Size = 5,386,000 \| Hispanic = 0.979 \| Census Detail and SSA POB Match = 0.932 | | | | | | | | | | |
| H1P | Hispanic, White, Census Detail/SSA POB Match | 0.516 | 0.516 | 9.360 | 0.000 | 0.000 | 0.000 | 9.951 | 0.000 | 0.000 | 0.000 |
| H6P | Hispanic, Some Other Race, Census Detail/SSA POB Match | 0.347 | 0.863 | -0.046 | 0.015 | -0.046 | -0.043 | -0.036 | 0.001 | -0.055 | -0.052 |
| H8P | Hispanic, 2+ Races (SOR), Census Detail/SSA POB Match | 0.035 | 0.898 | -0.058 | -0.019 | -0.049 | -0.048 | -0.058 | -0.036 | -0.054 | -0.052 |
| H1N | Hispanic, White | 0.023 | 0.921 | 0.076 | 0.057 | 0.037 | 0.014 | -0.067 | -0.078 | 0.003 | 0.015 |
| H2P | Hispanic, Black, Census Detail/SSA POB Match | 0.021 | 0.942 | -0.062 | -0.144 | -0.039 | -0.138 | -0.401 | -0.423 | -0.105 | -0.138 |
| H6N | Hispanic, Some Other Race | 0.020 | 0.962 | -0.062 | -0.011 | -0.028 | -0.043 | -0.185 | -0.156 | -0.071 | -0.052 |
| N1N | Not-Hispanic, White | 0.016 | 0.978 | 0.407 | 0.161 | 0.105 | 0.181 | 0.407 | 0.270 | 0.166 | 0.249 |
| XXX | All Other Responses | 0.009 | 0.987 | 0.239 | 0.169 | 0.087 | 0.047 | 0.053 | 0.010 | 0.014 | 0.040 |
| H3P | Hispanic, AIAN, Census Detail/SSA POB Match | 0.009 | 0.995 | -0.061 | -0.007 | -0.053 | -0.068 | -0.084 | -0.055 | -0.071 | -0.073 |
| H7P | Hispanic, 2+ Races (not SOR), Census Detail/SSA POB Match | 0.003 | 0.998 | 0.196 | 0.158 | 0.055 | -0.014 | 0.125 | 0.105 | 0.030 | -0.002 |
| N2N | Not-Hispanic, Black | 0.002 | 1.000 | 0.494 | 0.299 | 0.086 | 0.005 | -0.082 | -0.150 | -0.105 | -0.010 |
| MENA | Middle East and North Africa \| Sample Size = 510,000 \| Hispanic = 0.007 \| Census Detail and SSA POB Match = 0.247 | | | | | | | | | | |
| N1N | Not-Hispanic, White | 0.460 | 0.460 | 9.601 | 0.000 | 0.000 | 0.000 | 10.380 | 0.000 | 0.000 | 0.000 |
| N6N | Not-Hispanic, Some Other Race | 0.217 | 0.678 | -0.305 | -0.200 | -0.144 | -0.072 | -0.536 | -0.400 | -0.209 | -0.123 |
| N6P | Not-Hispanic, Some Other Race, Census Detail/SSA POB Match | 0.196 | 0.874 | 0.028 | 0.109 | 0.006 | -0.009 | -0.145 | -0.066 | -0.071 | -0.062 |
| N8P | Not-Hispanic, 2+ Races (SOR), Census Detail/SSA POB Match | 0.051 | 0.925 | 0.045 | 0.143 | 0.037 | -0.019 | -0.068 | -0.024 | -0.058 | -0.071 |
| N8N | Not-Hispanic, 2+ Races (SOR) | 0.030 | 0.955 | -0.402 | -0.281 | -0.176 | -0.121 | -0.533 | -0.408 | -0.206 | -0.155 |
| N4N | Not-Hispanic, Asian | 0.023 | 0.978 | 0.300 | 0.356 | 0.105 | -0.003 | 0.067 | 0.123 | -0.046 | -0.065 |
| N2N | Not-Hispanic, Black | 0.011 | 0.988 | 0.067 | 0.132 | -0.111 | -0.243 | -0.503 | -0.330 | -0.271 | -0.304 |
| XXX | All Other Responses | 0.007 | 0.996 | -0.082 | -0.074 | -0.162 | -0.174 | -0.478 | -0.380 | -0.276 | -0.225 |
| H1N | Hispanic, White | 0.004 | 1.000 | 0.266 | 0.208 | 0.031 | -0.146 | -0.047 | -0.043 | -0.085 | -0.153 |
| OFB | Other Foreign Born \| Sample Size = 8,272,000 \| Hispanic = 0.012 | | | | | | | | | | |
| N4N | Not-Hispanic, Asian | 0.477 | 0.477 | 9.974 | 0.000 | 0.000 | 0.000 | 10.530 | 0.000 | 0.000 | 0.000 |
| N1N | Not-Hispanic, White | 0.296 | 0.773 | -0.085 | -0.095 | -0.044 | 0.029 | 0.024 | 0.070 | 0.124 | 0.128 |
| N2N | Not-Hispanic, Black | 0.131 | 0.904 | -0.002 | 0.113 | -0.086 | -0.198 | -0.455 | -0.248 | -0.169 | -0.217 |
| N6N | Not-Hispanic, Some Other Race | 0.039 | 0.943 | -0.272 | -0.117 | -0.154 | -0.156 | -0.560 | -0.331 | -0.159 | -0.147 |
| N7N | Not-Hispanic, 2+ Races (not SOR) | 0.022 | 0.966 | -0.130 | -0.179 | -0.100 | -0.059 | -0.256 | -0.197 | -0.101 | -0.062 |
| N8N | Not-Hispanic, 2+ Races (SOR) | 0.013 | 0.979 | -0.204 | -0.141 | -0.108 | -0.081 | -0.257 | -0.152 | -0.065 | -0.054 |
| N5N | Not Hispanic, Hawaiian/Pacific Islander | 0.008 | 0.987 | -0.560 | -0.495 | -0.264 | -0.122 | -0.681 | -0.460 | -0.271 | -0.126 |
| XXX | All Other Responses | 0.007 | 0.994 | -0.288 | -0.276 | -0.177 | -0.198 | -0.505 | -0.347 | -0.168 | -0.183 |
| H1N | Hispanic, White | 0.006 | 1.000 | -0.211 | -0.227 | -0.114 | -0.134 | -0.248 | -0.129 | -0.035 | -0.089 |
| USA | USA Born \| Sample Size = 73,800,000 \| Hispanic = 0.068 | | | | | | | | | | |
| N1N | Not-Hispanic, White | 0.779 | 0.779 | 9.849 | 0.000 | 0.000 | 0.000 | 10.380 | 0.000 | 0.000 | 0.000 |
| N2N | Not-Hispanic, Black | 0.119 | 0.898 | -0.225 | -0.076 | -0.076 | -0.176 | -0.835 | -0.664 | -0.270 | -0.198 |
| H1N | Hispanic, White | 0.042 | 0.941 | -0.077 | 0.065 | -0.004 | -0.137 | -0.262 | -0.112 | -0.087 | -0.160 |
| H6N | Hispanic, Some Other Race | 0.018 | 0.959 | -0.238 | -0.021 | -0.067 | -0.191 | -0.434 | -0.203 | -0.147 | -0.211 |
| N7N | Not-Hispanic, 2+ Races (not SOR) | 0.010 | 0.969 | -0.223 | -0.139 | 0.009 | -0.051 | -0.354 | -0.271 | -0.084 | -0.071 |
| N4N | Not-Hispanic, Asian | 0.010 | 0.979 | 0.560 | 0.530 | 0.299 | 0.111 | 0.256 | 0.231 | 0.136 | 0.044 |
| XXX | All Other Responses | 0.009 | 0.988 | -0.289 | -0.124 | -0.029 | -0.171 | -0.532 | -0.360 | -0.146 | -0.182 |
| N3N | Not-Hispanic, American Indian/Alaska Native | 0.008 | 0.995 | -0.534 | -0.371 | -0.100 | -0.096 | -0.825 | -0.655 | -0.240 | -0.102 |
| N6N | Not-Hispanic, Some Other Race | 0.005 | 1.000 | -0.324 | -0.244 | -0.051 | -0.058 | -0.453 | -0.384 | -0.101 | -0.056 |

Notes: This table uses Sample 1, which consists of 87.97 million workers age 25 to 55 (inclusive) in 2010. All workers must have at least one year with positive earnings during the six year period from 2010 to 2015. Zero earning years are included in the earnings measures. Place of birth is determined using the Census Numident. The Hispanic place of birth group consists of the set of countries where at least 50% of foreign-born persons in Sample 1 report Hispanic on the 2010 Census. The full list of Hispanic and MENA countries can be found in Table 3. The race measure consists of eight categories collated from the 2010 Census: White, Black, American Indian/Alaska Native (AIAN), Asian, Native Hawaiian/Pacific Islander, Some Other Race (SOR), 2+ Races (not SOR), and 2+ Races (SOR). The ethnicity variable is binary; Hispanic/Latino or Not Hispanic/Latino. A "Census Detail/SSA POB Match" occurs when the 2010 Census respondent reported their Census Numident POB on the 2010 Race or Ethnicity questions using the detailed coding (the detailed Decennial POB coding is not available for 8 MENA countries as shown in Table 3). For both females and males we report four measures of mean log real long-term earnings. The first row of the Mean Log Earnings column for each POB group shows the mean log earnings of the largest Ethnicity-Race-Census Detail/SSA POB Match category, while the other rows report the difference in mean log earnings relative to the largest category. The next three columns show adjusted log real long-term earnings from linear regression models controlling for various person and job characteristics. The PC model includes controls for education and quadratic functions of age and years in the U.S. in 2010. The PCW model adds controls for the dominant industry/region and the number of full and partial years worked. The PCWH model adds a quadratic function of average annual hours worked. Standard errors have not been released yet for this table. Some sub-groups have fewer than 1,000 persons. Even for small sub-groups, standard errors for adjusted log earnings differentials rarely exceed 0.025 log points.

## Table 3 – Mean Log Long-Term Average Annual Earnings by Sex, Detailed Place of Birth, Ethnicity, Race, and Detail Census/SSA Place of Birth Match

| | | | | Females | | Males | |
|---|---|---|---|---|---|---|---|
| Row Label | Ethnicity - Race - SSA Census Detail/SSA POB Match | Propor-tion | Cumula-tive Pro-portion | Mean Log Earnings | Adjusted Log Earnings PCWH | Mean Log Earnings | Adjusted Log Earnings PCWH |
| ARG | Argentina \| Sample Size = 56,500 \| Hispanic = 0.903 \| Census Detail/SSA POB Match = 0.858 | | | | | | |
| H1P | Hispanic, White, Census Detail/SSA POB Match | 0.762 | 0.762 | 9.759 | 0.000 | 10.410 | 0.000 |
| N1N | Not Hispanic, White | 0.080 | 0.842 | 0.169 | 0.083 | 0.124 | 0.091 |
| H6P | Hispanic, Some Other Race, Census Detail/SSA POB Match | 0.069 | 0.911 | -0.207 | -0.105 | -0.459 | -0.153 |
| H1N | Hispanic, White | 0.034 | 0.945 | -0.093 | -0.010 | -0.228 | -0.047 |
| H8P | Hispanic, 2+ Races (SOR), Census Detail/SSA POB Match | 0.015 | 0.960 | -0.222 | -0.039 | -0.468 | -0.127 |
| N6N | Not Hispanic, Some Other Race | 0.012 | 0.972 | 0.081 | -0.045 | -0.682 | -0.072 |
| N4N | Not Hispanic, Asian | 0.010 | 0.981 | 0.347 | 0.088 | 0.200 | 0.023 |
| H6N | Hispanic, Some Other Race | 0.008 | 0.989 | -0.360 | -0.106 | -0.537 | -0.168 |
| XXX | All Other Responses | 0.008 | 0.997 | -0.032 | -0.022 | -0.357 | -0.178 |
| H4P | Hispanic, Asian , Census Detail/SSA POB Match | 0.003 | 1.000 | -0.049 | -0.033 | -0.193 | -0.036 |
| BOL | Bolivia \| Sample Size = 23,500 \| Hispanic = 0.936 \| Census Detail/SSA POB Match = 0.851 | | | | | | |
| H1P | Hispanic, White, Census Detail/SSA POB Match | 0.543 | 0.543 | 9.603 | 0.000 | 10.180 | 0.000 |
| H6P | Hispanic, Some Other Race, Census Detail/SSA POB Match | 0.243 | 0.787 | -0.084 | -0.036 | -0.124 | -0.052 |
| N1N | Not Hispanic, White | 0.052 | 0.839 | 0.309 | 0.156 | 0.285 | 0.160 |
| H8P | Hispanic, 2+ Races (SOR), Census Detail/SSA POB Match | 0.048 | 0.887 | -0.040 | -0.060 | -0.173 | -0.052 |
| H1N | Hispanic, White | 0.043 | 0.930 | 0.012 | -0.012 | -0.098 | 0.041 |
| H6N | Hispanic, Some Other Race | 0.024 | 0.954 | 0.102 | -0.054 | -0.237 | -0.057 |
| XXX | All Other Responses | 0.017 | 0.972 | 0.116 | 0.001 | -0.182 | -0.086 |
| H3P | Hispanic, AIAN, Census Detail/SSA POB Match | 0.013 | 0.985 | 0.075 | -0.005 | -0.135 | -0.051 |
| N4N | Not Hispanic, Asian | 0.009 | 0.993 | 0.158 | 0.149 | 0.152 | 0.134 |
| H7P | Hispanic, 2+ Races (not SOR), Census Detail/SSA POB Match | 0.007 | 1.000 | 0.421 | -0.002 | 0.033 | -0.057 |
| CHI | Chile \| Sample Size = 27,000 \| Hispanic = 0.926 \| Census Detail/SSA POB Match = 0.852 | | | | | | |
| H1P | Hispanic, White, Census Detail/SSA POB Match | 0.653 | 0.653 | 9.627 | 0.000 | 10.260 | 0.000 |
| H6P | Hispanic, Some Other Race, Census Detail/SSA POB Match | 0.160 | 0.813 | -0.154 | -0.058 | -0.272 | -0.105 |
| N1N | Not Hispanic, White | 0.052 | 0.866 | 0.161 | 0.113 | 0.314 | 0.193 |
| H1N | Hispanic, White | 0.041 | 0.907 | -0.056 | 0.030 | -0.157 | -0.029 |
| H8P | Hispanic, 2+ Races (SOR), Census Detail/SSA POB Match | 0.032 | 0.938 | -0.145 | -0.070 | -0.260 | -0.088 |
| N4N | Not Hispanic, Asian | 0.024 | 0.963 | 0.570 | 0.249 | 0.391 | 0.224 |
| H6N | Hispanic, Some Other Race | 0.021 | 0.983 | -0.196 | -0.098 | -0.383 | -0.116 |
| XXX | All Other Responses | 0.013 | 0.996 | -0.002 | -0.125 | -0.404 | -0.052 |
| H7P | Hispanic, 2+ Races (not SOR), Census Detail/SSA POB Match | 0.004 | 1.000 | 0.161 | -0.117 | -0.010 | -0.128 |
| COL | Columbia \| Sample Size = 243,000 \| Hispanic = 0.979 \| Census Detail/SSA POB Match = 0.926 | | | | | | |
| H1P | Hispanic, White, Census Detail/SSA POB Match | 0.692 | 0.692 | 9.528 | 0.000 | 10.130 | 0.000 |
| H6P | Hispanic, Some Other Race, Census Detail/SSA POB Match | 0.181 | 0.873 | -0.039 | -0.048 | -0.087 | -0.071 |
| H1N | Hispanic, White | 0.034 | 0.907 | -0.056 | 0.006 | -0.110 | -0.003 |
| H8P | Hispanic, 2+ Races (SOR), Census Detail/SSA POB Match | 0.034 | 0.941 | -0.082 | -0.056 | -0.110 | -0.066 |
| N1N | Not Hispanic, White | 0.015 | 0.956 | 0.298 | 0.142 | 0.344 | 0.179 |
| H2P | Hispanic, Black, Census Detail/SSA POB Match | 0.014 | 0.970 | -0.017 | -0.159 | -0.175 | -0.158 |
| H6N | Hispanic, Some Other Race | 0.014 | 0.984 | -0.098 | -0.046 | -0.204 | -0.073 |
| XXX | All Other Responses | 0.010 | 0.994 | -0.008 | -0.020 | -0.167 | -0.056 |
| H3P | Hispanic, AIAN, Census Detail/SSA POB Match | 0.004 | 0.997 | -0.113 | -0.060 | -0.347 | -0.104 |
| N4N | Not Hispanic, Asian | 0.003 | 1.000 | 0.596 | 0.071 | 0.326 | 0.106 |
| COS | Costa Rica \| Sample Size = 27,000 \| Hispanic = 0.926 \| Census Detail/SSA POB Match = 0.833 | | | | | | |
| H1P | Hispanic, White, Census Detail/SSA POB Match | 0.556 | 0.556 | 9.469 | 0.000 | 10.100 | 0.000 |
| H6P | Hispanic, Some Other Race, Census Detail/SSA POB Match | 0.185 | 0.741 | -0.121 | -0.062 | -0.180 | -0.093 |
| N1N | Not Hispanic, White | 0.059 | 0.800 | 0.282 | 0.153 | 0.367 | 0.202 |
| H1N | Hispanic, White | 0.052 | 0.852 | 0.018 | -0.018 | -0.200 | 0.012 |
| H2P | Hispanic, Black, Census Detail/SSA POB Match | 0.048 | 0.900 | 0.673 | -0.063 | 0.160 | -0.136 |
| H8P | Hispanic, 2+ Races (SOR), Census Detail/SSA POB Match | 0.041 | 0.941 | -0.013 | -0.062 | -0.176 | -0.087 |
| H6N | Hispanic, Some Other Race | 0.024 | 0.965 | -0.115 | -0.074 | -0.342 | -0.093 |
| XXX | All Other Responses | 0.019 | 0.983 | 0.289 | -0.020 | -0.340 | -0.042 |
| N2N | Not Hispanic, Black | 0.011 | 0.994 | 0.577 | 0.029 | -0.221 | -0.150 |
| N4N | Not Hispanic, Asian | 0.006 | 1.000 | 0.960 | 0.258 | 0.736 | 0.295 |
| CUB | Cuba \| Sample Size = 348,000 \| Hispanic = 0.989 \| Census Detail/SSA POB Match = 0.974 | | | | | | |
| H1P | Hispanic, White, Census Detail/SSA POB Match | 0.871 | 0.871 | 9.498 | 0.000 | 9.788 | 0.000 |
| H6P | Hispanic, Some Other Race, Census Detail/SSA POB Match | 0.050 | 0.921 | -0.142 | -0.049 | -0.225 | -0.068 |
| H2P | Hispanic, Black, Census Detail/SSA POB Match | 0.036 | 0.957 | -0.234 | -0.162 | -0.413 | -0.144 |
| H8P | Hispanic, 2+ Races (SOR), Census Detail/SSA POB Match | 0.013 | 0.970 | -0.166 | -0.063 | -0.310 | -0.078 |
| H1N | Hispanic, White | 0.010 | 0.980 | 0.136 | 0.025 | -0.148 | -0.004 |
| N1N | Not Hispanic, White | 0.010 | 0.991 | 0.392 | 0.097 | 0.487 | 0.092 |
| XXX | All Other Responses | 0.009 | 1.000 | 0.151 | -0.002 | -0.139 | -0.038 |
| DR | Dominican Republic \| Sample Size = 305,000 \| Hispanic = 0.990 \| Census Detail/SSA POB Match = 0.928 | | | | | | |
| H6P | Hispanic, Some Other Race, Census Detail/SSA POB Match | 0.439 | 0.439 | 9.327 | 0.000 | 9.726 | 0.000 |
| H1P | Hispanic, White, Census Detail/SSA POB Match | 0.285 | 0.725 | 0.071 | 0.045 | 0.125 | 0.048 |
| H2P | Hispanic, Black, Census Detail/SSA POB Match | 0.116 | 0.841 | 0.065 | -0.077 | 0.047 | -0.055 |
| H8P | Hispanic, 2+ Races (SOR), Census Detail/SSA POB Match | 0.071 | 0.912 | -0.020 | -0.018 | -0.020 | -0.009 |
| H6N | Hispanic, Some Other Race | 0.025 | 0.937 | -0.047 | -0.008 | -0.102 | 0.014 |
| H1N | Hispanic, White | 0.025 | 0.962 | 0.052 | 0.041 | -0.044 | 0.032 |
| XXX | All Other Responses | 0.013 | 0.975 | 0.013 | 0.044 | -0.004 | 0.017 |
| H3P | Hispanic, AIAN, Census Detail/SSA POB Match | 0.012 | 0.987 | -0.123 | -0.037 | -0.199 | -0.010 |
| H2N | Hispanic, Black | 0.007 | 0.993 | 0.005 | -0.074 | 0.033 | -0.050 |
| N1N | Not Hispanic, White | 0.004 | 0.997 | 0.191 | 0.214 | 0.410 | 0.213 |
| N2N | Not Hispanic, Black | 0.003 | 1.000 | 0.494 | 0.094 | 0.198 | 0.077 |
| ECU | Ecuador \| Sample Size = 114,000 \| Hispanic = 0.991 \| Census Detail/SSA POB Match = 0.921 | | | | | | |
| H1P | Hispanic, White, Census Detail/SSA POB Match | 0.505 | 0.505 | 9.542 | 0.000 | 10.100 | 0.000 |
| H6P | Hispanic, Some Other Race, Census Detail/SSA POB Match | 0.334 | 0.838 | -0.091 | -0.037 | -0.112 | -0.045 |
| H8P | Hispanic, 2+ Races (SOR), Census Detail/SSA POB Match | 0.058 | 0.896 | -0.150 | -0.045 | -0.134 | -0.050 |
| H1N | Hispanic, White | 0.037 | 0.933 | 0.020 | 0.005 | -0.090 | -0.014 |
| H6N | Hispanic, Some Other Race | 0.027 | 0.960 | -0.101 | -0.037 | -0.069 | -0.055 |
| N1N | Not Hispanic, White | 0.011 | 0.972 | 0.300 | 0.149 | 0.284 | 0.166 |
| H3P | Hispanic, AIAN, Census Detail/SSA POB Match | 0.009 | 0.981 | 0.077 | -0.063 | -0.266 | -0.087 |
| XXX | All Other Responses | 0.007 | 0.988 | 0.073 | -0.011 | -0.075 | 0.019 |
| H2P | Hispanic, Black, Census Detail/SSA POB Match | 0.005 | 0.993 | -0.165 | -0.143 | -0.273 | -0.153 |
| H7P | Hispanic, 2+ Races (not SOR), Census Detail/SSA POB Match | 0.004 | 0.997 | 0.209 | 0.022 | 0.124 | -0.001 |
| H8N | Hispanic, 2+ Races (SOR) | 0.003 | 1.000 | -0.173 | -0.037 | -0.056 | -0.056 |
| ELS | El Salvador \| Sample Size = 435,000 \| Hispanic = 0.993 \| Census Detail/SSA POB Match = 0.887 | | | | | | |
| H6P | Hispanic, Some Other Race, Census Detail/SSA POB Match | 0.465 | 0.465 | 9.423 | 0.000 | 9.927 | 0.000 |
| H1P | Hispanic, White, Census Detail/SSA POB Match | 0.350 | 0.815 | 0.032 | 0.026 | 0.053 | 0.025 |
| H8P | Hispanic, 2+ Races (SOR), Census Detail/SSA POB Match | 0.056 | 0.871 | -0.050 | -0.004 | 0.005 | 0.001 |
| H6N | Hispanic, Some Other Race | 0.055 | 0.926 | -0.070 | 0.000 | -0.120 | -0.004 |
| H1N | Hispanic, White | 0.045 | 0.971 | -0.010 | 0.021 | -0.130 | 0.015 |
| H3P | Hispanic, AIAN, Census Detail/SSA POB Match | 0.009 | 0.980 | 0.056 | -0.017 | 0.057 | -0.003 |
| XXX | All Other Responses | 0.009 | 0.989 | 0.059 | 0.018 | -0.057 | 0.012 |
| N1N | Not Hispanic, White | 0.004 | 0.993 | 0.117 | 0.170 | -0.013 | 0.170 |
| H8N | Hispanic, 2+ Races (SOR) | 0.004 | 0.997 | -0.082 | -0.002 | -0.174 | 0.000 |
| H2P | Hispanic, Black, Census Detail/SSA POB Match | 0.003 | 1.000 | 0.021 | -0.084 | -0.025 | -0.082 |
| GUA | Guatemala \| Sample Size = 159,000 \| Hispanic = 0.981 \| Census Detail/SSA POB Match = 0.862 | | | | | | |
| H6P | Hispanic, Some Other Race, Census Detail/SSA POB Match | 0.417 | 0.417 | 9.297 | 0.000 | 9.881 | 0.000 |
| H1P | Hispanic, White, Census Detail/SSA POB Match | 0.355 | 0.772 | 0.097 | 0.027 | 0.079 | 0.042 |
| H6N | Hispanic, Some Other Race | 0.061 | 0.833 | -0.078 | -0.006 | -0.175 | -0.011 |
| H8P | Hispanic, 2+ Races (SOR), Census Detail/SSA POB Match | 0.054 | 0.887 | -0.017 | -0.011 | -0.053 | 0.013 |
| H1N | Hispanic, White | 0.049 | 0.936 | 0.036 | 0.016 | -0.138 | 0.013 |
| H3P | Hispanic, AIAN, Census Detail/SSA POB Match | 0.019 | 0.956 | -0.007 | -0.016 | -0.064 | -0.025 |
| XXX | All Other Responses | 0.019 | 0.975 | 0.113 | 0.017 | -0.049 | 0.010 |
| N1N | Not Hispanic, White | 0.010 | 0.985 | 0.400 | 0.191 | 0.362 | 0.273 |
| H7P | Hispanic, 2+ Races (not SOR), Census Detail/SSA POB Match | 0.005 | 0.990 | 0.121 | 0.013 | 0.162 | 0.022 |
| H2P | Hispanic, Black, Census Detail/SSA POB Match | 0.005 | 0.995 | 0.239 | -0.145 | 0.060 | -0.091 |
| H8N | Hispanic, 2+ Races (SOR) | 0.005 | 1.000 | 0.023 | 0.002 | -0.168 | -0.008 |
| HON | Honduras \| Sample Size = 118,000 \| Hispanic = 0.975 \| Census Detail/SSA POB Match = 0.873 | | | | | | |
| H1P | Hispanic, White, Census Detail/SSA POB Match | 0.394 | 0.394 | 9.274 | 0.000 | 9.778 | 0.000 |
| H6P | Hispanic, Some Other Race, Census Detail/SSA POB Match | 0.356 | 0.750 | -0.012 | -0.036 | -0.012 | -0.039 |
| H8P | Hispanic, 2+ Races (SOR), Census Detail/SSA POB Match | 0.062 | 0.812 | -0.060 | -0.051 | -0.014 | -0.031 |
| H6N | Hispanic, Some Other Race | 0.049 | 0.861 | -0.095 | -0.046 | -0.141 | -0.056 |
| H1N | Hispanic, White | 0.045 | 0.906 | -0.017 | 0.000 | -0.060 | -0.015 |
| H2P | Hispanic, Black, Census Detail/SSA POB Match | 0.035 | 0.941 | 0.339 | -0.125 | 0.267 | -0.105 |
| XXX | All Other Responses | 0.019 | 0.959 | 0.292 | 0.023 | 0.164 | 0.051 |
| H3P | Hispanic, AIAN, Census Detail/SSA POB Match | 0.014 | 0.973 | 0.102 | -0.062 | 0.046 | -0.042 |
| N1N | Not Hispanic, White | 0.014 | 0.986 | 0.393 | 0.189 | 0.544 | 0.285 |
| H7P | Hispanic, 2+ Races (not SOR), Census Detail/SSA POB Match | 0.007 | 0.993 | 0.143 | -0.097 | 0.167 | 0.001 |
| N2N | Not Hispanic, Black | 0.007 | 1.000 | 0.485 | -0.013 | 0.220 | -0.007 |
| MEX | Mexico \| Sample Size = 2,582,000 \| Hispanic = 0.993 \| Census Detail/SSA POB Match = 0.966 | | | | | | |
| H1P | Hispanic, White, Census Detail/SSA POB Match | 0.500 | 0.500 | 9.266 | 0.000 | 10.020 | 0.000 |
| H6P | Hispanic, Some Other Race, Census Detail/SSA POB Match | 0.422 | 0.922 | -0.002 | -0.026 | -0.063 | -0.033 |
| H8P | Hispanic, 2+ Races (SOR), Census Detail/SSA POB Match | 0.028 | 0.950 | -0.072 | -0.024 | -0.047 | -0.035 |
| H6N | Hispanic, Some Other Race | 0.013 | 0.962 | -0.071 | -0.033 | -0.183 | -0.041 |
| H1N | Hispanic, White | 0.013 | 0.975 | -0.071 | -0.009 | -0.168 | -0.005 |
| H3P | Hispanic, AIAN, Census Detail/SSA POB Match | 0.010 | 0.985 | 0.021 | -0.051 | -0.064 | -0.048 |
| XXX | All Other Responses | 0.009 | 0.994 | 0.075 | -0.026 | -0.078 | -0.032 |
| N1N | Not Hispanic, White | 0.006 | 1.000 | 0.114 | 0.161 | 0.075 | 0.219 |
| NIC | Nicaragua \| Sample Size = 93,000 \| Hispanic = 0.984 \| Census Detail/SSA POB Match = 0.903 | | | | | | |
| H1P | Hispanic, White, Census Detail/SSA POB Match | 0.595 | 0.595 | 9.505 | 0.000 | 9.956 | 0.000 |
| H6P | Hispanic, Some Other Race, Census Detail/SSA POB Match | 0.243 | 0.838 | -0.041 | -0.038 | 0.002 | -0.038 |
| H1N | Hispanic, White | 0.045 | 0.883 | -0.002 | -0.003 | -0.063 | -0.019 |
| H8P | Hispanic, 2+ Races (SOR), Census Detail/SSA POB Match | 0.043 | 0.926 | -0.024 | -0.044 | -0.096 | -0.038 |
| H6N | Hispanic, Some Other Race | 0.030 | 0.957 | -0.090 | -0.055 | -0.176 | -0.052 |
| H2P | Hispanic, Black, Census Detail/SSA POB Match | 0.013 | 0.970 | 0.256 | -0.087 | -0.082 | -0.104 |
| XXX | All Other Responses | 0.011 | 0.981 | 0.362 | 0.094 | 0.179 | 0.059 |
| N1N | Not Hispanic, White | 0.007 | 0.988 | 0.227 | 0.172 | 0.336 | 0.191 |
| H3P | Hispanic, AIAN, Census Detail/SSA POB Match | 0.006 | 0.994 | -0.193 | -0.028 | -0.112 | -0.090 |
| H7P | Hispanic, 2+ Races (not SOR), Census Detail/SSA POB Match | 0.003 | 0.997 | 0.264 | 0.043 | 0.324 | -0.012 |
| H8N | Hispanic, 2+ Races (SOR) | 0.003 | 1.000 | 0.028 | 0.064 | -0.142 | -0.058 |
| PAN | Panama \| Sample Size = 49,000 \| Hispanic = 0.724 \| Census Detail/SSA POB Match = 0.633 | | | | | | |
| H1P | Hispanic, White, Census Detail/SSA POB Match | 0.213 | 0.213 | 9.641 | 0.000 | 10.350 | 0.000 |
| H2P | Hispanic, Black, Census Detail/SSA POB Match | 0.199 | 0.413 | 0.364 | -0.156 | -0.155 | -0.193 |
| N1N | Not Hispanic, White | 0.195 | 0.608 | 0.211 | 0.109 | 0.089 | 0.101 |
| H6P | Hispanic, Some Other Race, Census Detail/SSA POB Match | 0.159 | 0.766 | -0.063 | -0.075 | -0.331 | -0.130 |
| N2N | Not Hispanic, Black | 0.057 | 0.823 | 0.235 | -0.090 | -0.465 | -0.130 |
| H1N | Hispanic, White | 0.049 | 0.872 | 0.209 | 0.029 | -0.075 | -0.014 |
| H8P | Hispanic, 2+ Races (SOR), Census Detail/SSA POB Match | 0.045 | 0.917 | -0.003 | -0.128 | -0.381 | -0.189 |
| XXX | All Other Responses | 0.035 | 0.951 | 0.144 | 0.004 | -0.069 | 0.005 |
| H6N | Hispanic, Some Other Race | 0.024 | 0.976 | -0.241 | -0.090 | -0.435 | -0.108 |
| N4N | Not Hispanic, Asian | 0.012 | 0.988 | 0.467 | 0.183 | 0.301 | 0.180 |
| H2N | Not Hispanic, Black | 0.012 | 1.000 | -0.018 | -0.205 | -0.433 | -0.211 |
| PER | Peru \| Sample Size = 143,000 \| Hispanic = 0.972 \| Census Detail/SSA POB Match = 0.916 | | | | | | |
| H1P | Hispanic, White, Census Detail/SSA POB Match | 0.501 | 0.501 | 9.615 | 0.000 | 10.140 | 0.000 |
| H6P | Hispanic, Some Other Race, Census Detail/SSA POB Match | 0.322 | 0.823 | -0.053 | -0.042 | -0.094 | -0.067 |
| H8P | Hispanic, 2+ Races (SOR), Census Detail/SSA POB Match | 0.050 | 0.873 | -0.071 | -0.038 | -0.086 | -0.054 |
| H1N | Hispanic, White | 0.032 | 0.905 | -0.022 | 0.002 | -0.072 | 0.000 |
| H6N | Hispanic, Some Other Race | 0.024 | 0.929 | -0.121 | -0.034 | -0.223 | -0.079 |
| N1N | Not Hispanic, White | 0.020 | 0.948 | 0.160 | 0.107 | 0.363 | 0.218 |
| H3P | Hispanic, AIAN, Census Detail/SSA POB Match | 0.013 | 0.961 | -0.104 | -0.066 | -0.152 | -0.096 |
| XXX | All Other Responses | 0.013 | 0.974 | 0.311 | 0.089 | 0.119 | 0.030 |
| H7P | Hispanic, 2+ Races (not SOR), Census Detail/SSA POB Match | 0.011 | 0.985 | 0.152 | -0.005 | 0.249 | 0.019 |
| H4P | Hispanic, Asian , Census Detail/SSA POB Match | 0.009 | 0.994 | 0.433 | 0.046 | 0.350 | 0.070 |
| H2P | Hispanic, Black, Census Detail/SSA POB Match | 0.006 | 1.000 | 0.005 | -0.181 | -0.203 | -0.155 |
| PR | Puerto Rico \| Sample Size = 530,000 \| Hispanic = 0.968 \| Census Detail/SSA POB Match = 0.934 | | | | | | |
| H1P | Hispanic, White, Census Detail/SSA POB Match | 0.597 | 0.597 | 9.137 | 0.000 | 9.478 | 0.000 |
| H6P | Hispanic, Some Other Race, Census Detail/SSA POB Match | 0.238 | 0.834 | 0.100 | -0.049 | 0.127 | -0.059 |
| H2P | Hispanic, Black, Census Detail/SSA POB Match | 0.065 | 0.900 | -0.360 | -0.134 | -0.438 | -0.131 |
| H8P | Hispanic, 2+ Races (SOR), Census Detail/SSA POB Match | 0.025 | 0.925 | 0.089 | -0.052 | 0.109 | -0.053 |
| N1N | Not Hispanic, White | 0.024 | 0.949 | 0.732 | 0.166 | 0.848 | 0.231 |
| H1N | Hispanic, White | 0.018 | 0.966 | 0.441 | 0.027 | 0.221 | 0.023 |
| H6N | Hispanic, Some Other Race | 0.013 | 0.979 | 0.208 | -0.051 | -0.119 | -0.052 |
| XXX | All Other Responses | 0.008 | 0.987 | 0.287 | -0.034 | 0.176 | -0.019 |
| H3P | Hispanic, AIAN, Census Detail/SSA POB Match | 0.007 | 0.994 | -0.111 | -0.062 | -0.128 | -0.095 |
| N4N | Not Hispanic, Asian | 0.003 | 0.997 | 1.373 | 0.185 | 1.344 | 0.181 |
| N2N | Not Hispanic, Black | 0.003 | 1.000 | 0.731 | -0.023 | 0.253 | 0.010 |
| SPA | Spain \| Sample Size = 44,000 \| Hispanic = 0.625 \| Census Detail/SSA POB Match = 0.489 | | | | | | |
| H1P | Hispanic, White, Census Detail/SSA POB Match | 0.454 | 0.454 | 10.030 | 0.000 | 10.790 | 0.000 |
| N1N | Not Hispanic, White | 0.317 | 0.771 | -0.143 | 0.058 | -0.346 | 0.017 |
| H1N | Hispanic, White | 0.113 | 0.884 | -0.092 | -0.007 | -0.223 | -0.003 |
| N2N | Not Hispanic, Black | 0.027 | 0.912 | -0.070 | -0.102 | -0.853 | -0.159 |
| H6P | Hispanic, Some Other Race, Census Detail/SSA POB Match | 0.027 | 0.939 | -0.176 | -0.106 | -0.366 | -0.115 |
| XXX | All Other Responses | 0.017 | 0.956 | -0.284 | -0.113 | -0.560 | -0.108 |
| H6N | Hispanic, Some Other Race | 0.014 | 0.969 | -0.256 | -0.067 | -0.488 | -0.066 |
| N4N | Not Hispanic, Asian | 0.010 | 0.980 | 0.193 | 0.159 | -0.162 | -0.048 |
| H8P | Hispanic, 2+ Races (SOR), Census Detail/SSA POB Match | 0.010 | 0.990 | -0.276 | -0.083 | -0.445 | -0.216 |
| N7N | Not Hispanic, 2+ Races (not SOR) | 0.007 | 0.997 | -0.193 | 0.016 | -0.496 | -0.012 |
| H8N | Hispanic, 2+ Races (SOR) | 0.003 | 1.000 | -0.147 | -0.081 | -0.641 | -0.129 |
| URU | Uruguay \| Sample Size = 11,500 \| Hispanic = 0.913 \| Census Detail/SSA POB Match = 0.843 | | | | | | |
| H1P | Hispanic, White, Census Detail/SSA POB Match | 0.752 | 0.752 | 9.611 | 0.000 | 10.100 | 0.000 |
| H6P | Hispanic, Some Other Race, Census Detail/SSA POB Match | 0.075 | 0.827 | -0.101 | -0.073 | -0.168 | -0.098 |
| N1N | Not Hispanic, White | 0.066 | 0.894 | 0.238 | 0.144 | 0.453 | 0.171 |
| H1N | Hispanic, White | 0.058 | 0.951 | 0.167 | 0.007 | -0.073 | -0.006 |
| H8P | Hispanic, 2+ Races (SOR), Census Detail/SSA POB Match | 0.022 | 0.973 | -0.202 | -0.029 | -0.334 | -0.115 |
| H6N | Hispanic, Some Other Race | 0.013 | 0.987 | -0.505 | -0.122 | -0.157 | -0.100 |
| XXX | All Other Responses | 0.013 | 1.000 | 0.082 | -0.023 | 0.097 | -0.033 |
| VEN | Venezuela \| Sample Size = 78,000 \| Hispanic = 0.891 \| Census Detail/SSA POB Match = 0.814 | | | | | | |
| H1P | Hispanic, White, Census Detail/SSA POB Match | 0.641 | 0.641 | 9.574 | 0.000 | 10.310 | 0.000 |
| H6P | Hispanic, Some Other Race, Census Detail/SSA POB Match | 0.126 | 0.766 | -0.055 | -0.071 | -0.116 | -0.112 |
| N1N | Not Hispanic, White | 0.077 | 0.843 | 0.406 | 0.116 | 0.332 | 0.109 |
| H1N | Hispanic, White | 0.056 | 0.899 | 0.202 | 0.023 | 0.018 | 0.011 |
| H8P | Hispanic, 2+ Races (SOR), Census Detail/SSA POB Match | 0.027 | 0.926 | -0.151 | -0.074 | -0.171 | -0.111 |
| N4N | Not Hispanic, Asian | 0.020 | 0.947 | 0.396 | 0.053 | 0.060 | -0.099 |
| H6N | Hispanic, Some Other Race | 0.017 | 0.963 | -0.013 | -0.110 | -0.273 | -0.143 |
| XXX | All Other Responses | 0.014 | 0.978 | 0.176 | -0.040 | -0.012 | -0.126 |
| H2P | Hispanic, Black, Census Detail/SSA POB Match | 0.013 | 0.990 | 0.060 | -0.172 | -0.335 | -0.232 |
| N6N | Not Hispanic, Some Other Race | 0.006 | 0.997 | -0.026 | 0.071 | -0.366 | -0.031 |
| N2N | Not Hispanic, Black | 0.003 | 1.000 | 0.322 | -0.120 | -0.315 | -0.175 |
| ALG | Algeria \| Sample Size = 7,500 \| Hispanic = 0.067 | | | | | | |
| N1N | Not Hispanic, White | 0.627 | 0.627 | 9.418 | 0.000 | 10.230 | 0.000 |
| N6N | Not Hispanic, Some Other Race | 0.253 | 0.880 | -0.360 | -0.074 | -0.366 | -0.022 |
| H1N | Hispanic, White | 0.053 | 0.933 | 0.643 | -0.070 | 0.190 | 0.048 |
| N8N | Not Hispanic, 2+ Races (SOR) | 0.027 | 0.960 | 0.084 | -0.034 | -0.118 | -0.066 |
| N2N | Not Hispanic, Black | 0.020 | 0.980 | 0.325 | -0.067 | -0.269 | -0.089 |
| XXX | All Other Responses | 0.020 | 1.000 | 0.638 | -0.099 | 0.301 | 0.034 |
| BAH | Bahrain \| Sample Size = 2,800 \| Hispanic = 0.014 | | | | | | |
| N4N | Not Hispanic, Asian | 0.473 | 0.473 | 9.808 | 0.000 | 10.580 | 0.000 |
| N1N | Not Hispanic, White | 0.218 | 0.691 | 0.299 | 0.044 | -0.028 | 0.024 |
| N2N | Not Hispanic, Black | 0.182 | 0.873 | 0.256 | -0.156 | -0.470 | -0.261 |
| XXX | All Other Responses | 0.073 | 0.945 | -0.004 | -0.032 | -0.193 | 0.012 |
| N6N | Not Hispanic, Some Other Race | 0.055 | 1.000 | -0.302 | -0.095 | -0.356 | -0.078 |
| EGY | Egypt \| Sample Size = 55,500 \| Hispanic = 0.005 \| Census Detail/SSA POB Match = 0.450 | | | | | | |
| N6P | Not Hispanic, Some Other Race, Census Detail/SSA POB Match | 0.432 | 0.432 | 9.501 | 0.000 | 10.140 | 0.000 |
| N1N | Not Hispanic, White | 0.369 | 0.802 | 0.137 | 0.051 | 0.262 | 0.080 |
| N6N | Not Hispanic, Some Other Race | 0.137 | 0.939 | -0.061 | 0.046 | 0.095 | 0.058 |
| N2N | Not Hispanic, Black | 0.020 | 0.959 | 0.039 | -0.111 | -0.344 | -0.112 |
| N8P | Not Hispanic, 2+ Races (SOR), Census Detail/SSA POB Match | 0.020 | 0.978 | -0.217 | -0.046 | -0.267 | -0.027 |
| N8N | Not Hispanic, 2+ Races (SOR) | 0.009 | 0.987 | -0.085 | 0.049 | -0.279 | -0.070 |
| N4N | Not Hispanic, Asian | 0.006 | 0.994 | 0.354 | 0.070 | 0.290 | 0.059 |
| XXX | All Other Responses | 0.006 | 1.000 | 0.235 | -0.165 | -0.134 | -0.145 |
| IRN | Iran \| Sample Size = 148,000 \| Hispanic = 0.002 \| Census Detail/SSA POB Match = 0.514 | | | | | | |
| N6P | Not Hispanic, Some Other Race, Census Detail/SSA POB Match | 0.380 | 0.380 | 9.817 | 0.000 | 10.480 | 0.000 |
| N1N | Not Hispanic, White | 0.319 | 0.699 | 0.118 | -0.014 | 0.115 | -0.001 |
| N8P | Not Hispanic, 2+ Races (SOR), Census Detail/SSA POB Match | 0.136 | 0.834 | -0.036 | -0.030 | -0.018 | -0.044 |
| N6N | Not Hispanic, Some Other Race | 0.136 | 0.970 | -0.143 | -0.135 | -0.626 | -0.163 |
| N8N | Not Hispanic, 2+ Races (SOR) | 0.015 | 0.985 | -0.175 | -0.184 | -0.614 | -0.168 |
| N4N | Not Hispanic, Asian | 0.009 | 0.994 | 0.092 | 0.029 | 0.026 | -0.039 |
| XXX | All Other Responses | 0.006 | 1.000 | 0.023 | -0.198 | -0.350 | -0.245 |
| IRQ | Iraq \| Sample Size = 50,000 \| Hispanic = 0.003 \| Census Detail/SSA POB Match = 0.102 | | | | | | |
| N6N | Not Hispanic, Some Other Race | 0.449 | 0.449 | 9.025 | 0.000 | 9.525 | 0.000 |
| N1N | Not Hispanic, White | 0.369 | 0.818 | 0.051 | 0.019 | 0.175 | 0.043 |
| N6P | Not Hispanic, Some Other Race, Census Detail/SSA POB Match | 0.068 | 0.886 | -0.535 | -0.011 | -0.384 | -0.040 |
| N8N | Not Hispanic, 2+ Races (SOR) | 0.056 | 0.942 | -0.131 | -0.060 | -0.050 | -0.050 |
| N8P | Not Hispanic, 2+ Races (SOR), Census Detail/SSA POB Match | 0.034 | 0.976 | -0.483 | -0.039 | -0.080 | -0.049 |
| N4N | Not Hispanic, Asian | 0.011 | 0.987 | 0.400 | 0.133 | 0.040 | 0.006 |
| N7N | Not Hispanic, 2+ Races | 0.009 | 0.996 | -0.441 | -0.052 | 0.053 | -0.070 |
| XXX | All Other Responses | 0.004 | 1.000 | -0.130 | -0.136 | -0.440 | -0.110 |
| ISR | Israel \| Sample Size = 59,000 \| Hispanic = 0.008 \| Census Detail/SSA POB Match = 0.051 | | | | | | |
| N1N | Not Hispanic, White | 0.796 | 0.796 | 9.616 | 0.000 | 10.440 | 0.000 |
| N6N | Not Hispanic, Some Other Race | 0.115 | 0.911 | -0.508 | -0.070 | -0.720 | -0.179 |
| N6P | Not Hispanic, Some Other Race, Census Detail/SSA POB Match | 0.044 | 0.955 | -0.417 | -0.106 | -0.613 | -0.119 |
| N8N | Not Hispanic, 2+ Races (SOR) | 0.020 | 0.975 | -0.607 | -0.152 | -0.728 | -0.225 |
| XXX | All Other Responses | 0.007 | 0.982 | -0.226 | -0.228 | -0.830 | -0.226 |
| H1N | Hispanic, White | 0.007 | 0.989 | 0.441 | -0.067 | 0.044 | -0.105 |
| N4N | Not Hispanic, Asian | 0.006 | 0.995 | -0.476 | -0.107 | -0.590 | -0.238 |
| N8P | Not Hispanic, 2+ Races (SOR), Census Detail/SSA POB Match | 0.005 | 1.000 | -0.165 | -0.176 | -0.359 | -0.153 |
| JOR | Jordan \| Sample Size = 30,000 \| Hispanic = 0.003 | | | | | | |
| N1N | Not Hispanic, White | 0.469 | 0.469 | 9.171 | 0.000 | 10.150 | 0.000 |
| N6N | Not Hispanic, Some Other Race | 0.419 | 0.888 | -0.151 | -0.016 | -0.363 | -0.052 |
| N8N | Not Hispanic, 2+ Races (SOR) | 0.090 | 0.978 | -0.150 | -0.104 | -0.226 | -0.083 |
| N4N | Not Hispanic, Asian | 0.013 | 0.992 | -0.223 | -0.055 | -0.489 | -0.074 |
| XXX | All Other Responses | 0.008 | 1.000 | 0.249 | -0.060 | -0.436 | -0.158 |
| KUW | Kuwait \| Sample Size = 12,500 \| Hispanic = 0.003 | | | | | | |
| N1N | Not Hispanic, White | 0.392 | 0.392 | 9.443 | 0.000 | 10.380 | 0.000 |
| N6N | Not Hispanic, Some Other Race | 0.337 | 0.729 | -0.068 | -0.015 | -0.304 | -0.071 |
| N4N | Not Hispanic, Asian | 0.188 | 0.918 | 0.752 | 0.039 | 0.517 | 0.051 |
| N8N | Not Hispanic, 2+ Races (SOR) | 0.067 | 0.984 | -0.324 | -0.051 | -0.132 | -0.070 |
| XXX | All Other Responses | 0.016 | 1.000 | -0.232 | -0.035 | -0.245 | -0.166 |
| LEB | Lebanon \| Sample Size = 53,500 \| Hispanic = 0.004 \| Census Detail/SSA POB Match = 0.148 | | | | | | |
| N1N | Not Hispanic, White | 0.583 | 0.583 | 9.647 | 0.000 | 10.620 | 0.000 |
| N6N | Not Hispanic, Some Other Race | 0.226 | 0.808 | -0.065 | -0.041 | -0.343 | -0.102 |
| N6P | Not Hispanic, Some Other Race, Census Detail/SSA POB Match | 0.117 | 0.925 | -0.338 | -0.072 | -0.452 | -0.123 |
| N8P | Not Hispanic, 2+ Races (SOR), Census Detail/SSA POB Match | 0.032 | 0.957 | -0.236 | -0.073 | -0.247 | -0.115 |
| N8N | Not Hispanic, 2+ Races (SOR) | 0.023 | 0.979 | -0.057 | -0.028 | -0.269 | -0.145 |
| N4N | Not Hispanic, Asian | 0.009 | 0.989 | 0.361 | -0.157 | -0.307 | -0.321 |
| XXX | All Other Responses | 0.007 | 0.995 | -0.140 | -0.157 | -0.629 | -0.180 |
| N2N | Not Hispanic, Black | 0.005 | 1.000 | 0.412 | -0.376 | -0.400 | -0.651 |
| LIB | Libya \| Sample Size = 5,400 \| Hispanic = 0.019 | | | | | | |
| N1N | Not Hispanic, White | 0.642 | 0.642 | 9.908 | 0.000 | 10.490 | 0.000 |
| N6N | Not Hispanic, Some Other Race | 0.174 | 0.817 | -0.830 | -0.029 | -0.488 | 0.022 |
| N4N | Not Hispanic, Asian | 0.092 | 0.908 | 0.437 | 0.190 | 0.346 | 0.075 |
| N2N | Not Hispanic, Black | 0.046 | 0.954 | -0.110 | -0.295 | -0.336 | -0.307 |
| XXX | All Other Responses | 0.046 | 1.000 | 0.128 | -0.120 | -0.084 | -0.083 |
| MOR | Morocco \| Sample Size = 35,500 \| Hispanic = 0.011 \| SSA POB Match = 0.197 | | | | | | |
| N1N | Not Hispanic, White | 0.536 | 0.536 | 9.453 | 0.000 | 10.160 | 0.000 |
| N6N | Not Hispanic, Some Other Race | 0.183 | 0.719 | -0.328 | -0.021 | -0.325 | -0.034 |
| N6P | Not Hispanic, Some Other Race, Census Detail/SSA POB Match | 0.181 | 0.900 | -0.285 | -0.031 | -0.373 | -0.064 |
| N2N | Not Hispanic, Black | 0.042 | 0.942 | -0.141 | -0.161 | -0.352 | -0.168 |
| N8N | Not Hispanic, 2+ Races (SOR) | 0.018 | 0.961 | -0.147 | -0.096 | -0.429 | -0.075 |
| N8P | Not Hispanic, 2+ Races (SOR), Census Detail/SSA POB Match | 0.016 | 0.976 | -0.307 | -0.071 | -0.530 | -0.128 |
| N4N | Not Hispanic, Asian | 0.008 | 0.984 | -0.217 | -0.005 | -0.348 | -0.079 |
| H1N | Hispanic, White | 0.008 | 0.993 | 0.498 | -0.017 | 0.379 | -0.026 |
| XXX | All Other Responses | 0.007 | 1.000 | -0.171 | -0.217 | -0.258 | -0.141 |
| SAU | Saudi Arabia \| Sample Size = 14,500 \| Hispanic = 0.059 | | | | | | |
| N1N | Not Hispanic, White | 0.486 | 0.486 | 9.853 | 0.000 | 10.660 | 0.000 |
| N4N | Not Hispanic, Asian | 0.162 | 0.648 | 0.096 | -0.027 | -0.010 | -0.093 |
| N6N | Not Hispanic, Some Other Race | 0.155 | 0.803 | -0.531 | -0.023 | -0.847 | -0.123 |
| N2N | Not Hispanic, Black | 0.070 | 0.873 | 0.050 | -0.294 | -0.648 | -0.414 |
| N8N | Not Hispanic, 2+ Races (SOR) | 0.042 | 0.915 | -0.419 | -0.171 | -0.706 | -0.085 |
| H1N | Hispanic, White | 0.032 | 0.947 | -0.113 | -0.297 | -0.404 | -0.321 |
| H6N | Hispanic, Some Other Race | 0.021 | 0.968 | -0.264 | -0.425 | -0.620 | -0.528 |
| N7N | Not Hispanic, 2+ Races | 0.021 | 0.989 | 0.082 | -0.046 | -0.406 | -0.278 |
| XXX | All Other Responses | 0.011 | 1.000 | -0.171 | -0.275 | -0.532 | -0.431 |
| SYR | Syria \| Sample Size = 24,000 \| Hispanic = 0.003 \| Census Detail/SSA POB Match = 0.096 | | | | | | |
| N1N | Not Hispanic, White | 0.561 | 0.561 | 9.319 | 0.000 | 10.520 | 0.000 |
| N6N | Not Hispanic, Some Other Race | 0.291 | 0.852 | -0.059 | -0.060 | -0.268 | -0.053 |
| N6P | Not Hispanic, Some Other Race, Census Detail/SSA POB Match | 0.067 | 0.919 | -0.357 | -0.050 | -0.503 | -0.107 |
| N8N | Not Hispanic, 2+ Races (SOR) | 0.037 | 0.956 | -0.112 | -0.139 | -0.149 | -0.045 |
| N8P | Not Hispanic, 2+ Races (SOR), Census Detail/SSA POB Match | 0.027 | 0.983 | -0.284 | -0.046 | -0.343 | -0.047 |
| N4N | Not Hispanic, Asian | 0.008 | 0.992 | 0.131 | -0.204 | -0.158 | -0.070 |
| XXX | All Other Responses | 0.008 | 1.000 | -0.005 | -0.199 | -0.279 | -0.118 |
| TUN | Tunisia \| Sample Size = 3,400 \| Hispanic = 0.006 | | | | | | |
| N1N | Not Hispanic, White | 0.559 | 0.559 | 9.688 | 0.000 | 10.260 | 0.000 |
| N6N | Not Hispanic, Some Other Race | 0.353 | 0.912 | -0.041 | -0.008 | -0.606 | -0.106 |
| XXX | All Other Responses | 0.088 | 1.000 | 0.208 | -0.053 | -0.203 | -0.171 |
| YEM | Yemen \| Sample Size = 9,500 \| Hispanic = 0.004 | | | | | | |
| N6N | Not Hispanic, Some Other Race | 0.561 | 0.561 | 8.160 | 0.000 | 9.396 | 0.000 |
| N1N | Not Hispanic, White | 0.212 | 0.772 | 0.324 | 0.106 | 0.218 | 0.032 |
| N8N | Not Hispanic, 2+ Races (SOR) | 0.148 | 0.921 | -0.210 | 0.038 | 0.118 | -0.029 |
| N4N | Not Hispanic, Asian | 0.058 | 0.979 | 1.262 | 0.048 | 0.318 | 0.048 |
| XXX | All Other Responses | 0.021 | 1.000 | 1.154 | -0.135 | 0.052 | -0.029 |

Notes: This table uses Sample 1, which consists of 87.97 million workers age 25 to 55 (inclusive) in 2010. All workers must have at least one year with positive earnings during the six year period from 2010 to 2015. Zero earning years are included in the earnings measures. Place of birth is determined using the Census Numident. The race measure consists of eight categories collated from the 2010 Census: White, Black, American Indian/Alaska Native (AIAN), Asian, Native Hawaiian/Pacific Islander, Some Other Race (SOR), 2+ Races (not SOR), and 2+ Races (SOR). The ethnicity variable is binary; Hispanic/Latino or Not Hispanic/Latino. A "Census Detail/SSA POB Match" occurs when the 2010 Census respondent reported their Census Numident POB on the 2010 Race or Ethnicity questions using the detailed coding(the detailed Decennial POB coding is not available for 8 MENA countries). For both females and males we report four measures of mean log real long-term earnings. The first row of the Mean Log Earnings column for each POB group shows the mean log earnings of the largest Ethnicity-Race-Census Detail/SSA POB Match category, while the other rows report the difference in mean log earnings relative to the largest category. The next three columns show adjusted earnings from linear regression models controlling for various person and job characteristics. The "PC" model includes controls for education and quadratic functions of age and years in the U.S. in 2010. The "PCW" model adds controls for the dominant industry/region and the number of full and partial years worked. The "PCWH" model adds a quadratic function of average annual hours worked. Standard errors have not been released yet for this table. Some sub-groups have fewer than 1,000 persons. Even for small sub-groups, standard errors for adjusted log earnings differentials rarely exceed 0.025 log points.

## Table 4 – Mean Log Long-Term Average Annual Earnings by Sex, Parent's Place of Birth, Ethnicity, Race, and Census Detail/SSA POB Match

| | | | | Females | | Males | |
|---|---|---|---|---|---|---|---|
| Row Label | Ethnicity - Race - Census Detail/SSA POB Match | Propor-tion | Cumula-tive Pro-portion | Mean Log Earnings | Adjusted Log Earnings PCWH | Mean Log Earnings | Adjusted Log Earnings PCWH |
| 1FBH | Single Foreign-Born Hispanic Parent \| Sample Size = 115,000 \| Hispanic = 0.948 \| Census Detail/SSA POB Match = 0.848 | | | | | | |
| H1P | Hispanic, White, Census Detail/SSA POB Match | 0.429 | 0.429 | 9.619 | 0.000 | 9.792 | 0.000 |
| H6P | Hispanic, Some Other Race, Census Detail/SSA POB Match | 0.355 | 0.785 | -0.072 | -0.030 | -0.027 | -0.020 |
| H1N | Hispanic, White | 0.049 | 0.834 | -0.059 | 0.004 | -0.149 | 0.002 |
| H6N | Hispanic, Some Other Race | 0.040 | 0.874 | -0.127 | -0.006 | -0.162 | -0.023 |
| N1N | Not Hispanic, White | 0.039 | 0.913 | -0.030 | 0.133 | -0.035 | 0.123 |
| H8P | Hispanic, 2+ Races (SOR), Census Detail/SSA POB Match | 0.027 | 0.940 | -0.073 | -0.025 | -0.128 | -0.022 |
| H2P | Hispanic, Black, Census Detail/SSA POB Match | 0.023 | 0.963 | 0.002 | -0.113 | -0.199 | -0.090 |
| XXX | All Other Responses | 0.016 | 0.980 | -0.055 | -0.005 | -0.124 | -0.026 |
| H3P | Hispanic, AIAN, Census Detail/SSA POB Match | 0.009 | 0.988 | -0.212 | -0.075 | -0.169 | -0.047 |
| N2N | Not Hispanic, Black | 0.008 | 0.996 | -0.298 | -0.031 | -0.470 | -0.037 |
| H8N | Hispanic, 2+ Races (SOR) | 0.004 | 1.000 | -0.078 | -0.014 | -0.314 | -0.021 |
| 2P1FBH | Two Parents, One Foreign-Born Hispanic, One Born USA \| Sample Size = 97,500 \| Hispanic = 0.805 \| Census Detail/SSA POB Match = 0.738 | | | | | | |
| H1P | Hispanic, White, Census Detail/SSA POB Match | 0.484 | 0.484 | 9.737 | 0.000 | 10.010 | 0.000 |
| H6P | Hispanic, Some Other Race, Census Detail/SSA POB Match | 0.190 | 0.674 | -0.166 | -0.044 | -0.117 | -0.048 |
| N1N | Not Hispanic, White | 0.175 | 0.849 | 0.037 | 0.128 | -0.016 | 0.123 |
| H1N | Hispanic, White | 0.043 | 0.892 | -0.133 | -0.001 | -0.088 | -0.012 |
| H8P | Hispanic, 2+ Races (SOR), Census Detail/SSA POB Match | 0.033 | 0.925 | -0.127 | -0.018 | -0.166 | -0.030 |
| H6N | Hispanic, Some Other Race | 0.022 | 0.947 | -0.286 | -0.038 | -0.240 | -0.053 |
| XXX | All Other Responses | 0.017 | 0.964 | -0.136 | 0.017 | -0.277 | -0.006 |
| H2P | Hispanic, Black, Census Detail/SSA POB Match | 0.012 | 0.976 | 0.062 | -0.129 | -0.236 | -0.140 |
| N2N | Not Hispanic, Black | 0.010 | 0.987 | -0.165 | -0.037 | -0.566 | -0.068 |
| H3P | Hispanic, AIAN, Census Detail/SSA POB Match | 0.008 | 0.995 | -0.225 | -0.061 | -0.261 | -0.031 |
| H7P | Hispanic, 2+ Races (not SOR), Census Detail/SSA POB Match | 0.005 | 1.000 | -0.088 | -0.008 | -0.091 | -0.041 |
| 2FBH | Two Foreign-Born Hispanic Parents \| Sample Size = 167,000 \| Hispanic = 0.976 \| Census Detail/SSA POB Match = 0.922 | | | | | | |
| H1P | Hispanic, White, Census Detail/SSA POB Match | 0.493 | 0.493 | 9.809 | 0.000 | 9.995 | 0.000 |
| H6P | Hispanic, Some Other Race, Census Detail/SSA POB Match | 0.388 | 0.880 | -0.093 | -0.029 | -0.027 | -0.018 |
| H1N | Hispanic, White | 0.029 | 0.909 | -0.065 | -0.008 | -0.224 | -0.014 |
| H6N | Hispanic, Some Other Race | 0.023 | 0.933 | -0.198 | -0.039 | -0.204 | -0.018 |
| H8P | Hispanic, 2+ Races (SOR), Census Detail/SSA POB Match | 0.021 | 0.954 | -0.138 | -0.019 | -0.047 | -0.009 |
| N1N | Not Hispanic, White | 0.017 | 0.971 | -0.013 | 0.106 | -0.147 | 0.086 |
| H2P | Hispanic, Black, Census Detail/SSA POB Match | 0.010 | 0.981 | -0.084 | -0.124 | -0.254 | -0.112 |
| XXX | All Other Responses | 0.010 | 0.990 | -0.121 | -0.004 | -0.164 | 0.006 |
| H3P | Hispanic, AIAN, Census Detail/SSA POB Match | 0.007 | 0.998 | -0.044 | -0.056 | -0.102 | -0.057 |
| H7P | Hispanic, 2+ Races (not SOR), Census Detail/SSA POB Match | 0.002 | 1.000 | 0.187 | -0.030 | -0.150 | -0.039 |
| 1MENA | Single Middle East and North Africa Parent \| Sample Size = 5,800 \| Hispanic = 0.026 \| Census Detail/SSA POB Match = 0.172 | | | | | | |
| N1N | Not Hispanic, White | 0.522 | 0.522 | 9.835 | 0.000 | 9.962 | 0.000 |
| N6N | Not Hispanic, Some Other Race | 0.226 | 0.748 | -0.060 | -0.042 | -0.208 | -0.063 |
| N6P | Not Hispanic, Some Other Race, Census Detail/SSA POB Match | 0.139 | 0.887 | 0.197 | 0.047 | 0.131 | 0.072 |
| XXX | All Other Responses | 0.113 | 1.000 | 0.060 | -0.073 | -0.168 | -0.069 |
| 2P1MENA | Two Parents, One MENA, One Born USA \| Sample Size = 12,500 \| Hispanic = 0.036 \| Census Detail/SSA POB Match = 0.072 | | | | | | |
| N1N | Not Hispanic, White | 0.803 | 0.803 | 9.992 | 0.000 | 10.180 | 0.000 |
| XXX | All Other Responses | 0.080 | 0.884 | -0.111 | -0.120 | -0.178 | -0.109 |
| N6N | Not Hispanic, Some Other Race | 0.064 | 0.948 | -0.033 | -0.020 | -0.148 | -0.044 |
| N6P | Not Hispanic, Some Other Race, Census Detail/SSA POB Match | 0.052 | 1.000 | -0.092 | 0.022 | -0.024 | 0.008 |
| 2MENA | Two Middle East and North Africa Parents \| Sample Size = 12,500 \| Hispanic = 0.000 \| Census Detail/SSA POB Match = 0.216 | | | | | | |
| N1N | Not Hispanic, White | 0.448 | 0.448 | 10.020 | 0.000 | 10.190 | 0.000 |
| N6N | Not Hispanic, Some Other Race | 0.304 | 0.752 | -0.150 | -0.061 | -0.205 | -0.050 |
| N6P | Not Hispanic, Some Other Race, Census Detail/SSA POB Match | 0.184 | 0.936 | 0.249 | 0.074 | 0.085 | 0.030 |
| XXX | All Other Responses | 0.036 | 0.972 | -0.154 | -0.085 | -0.075 | -0.008 |
| N8P | Not Hispanic, 2+ Races (SOR), Census Detail/SSA POB Match | 0.028 | 1.000 | 0.226 | -0.002 | 0.176 | 0.120 |
| 1OFB | Single Other Foreign-Born Parent \| Sample Size = 100,000 \| Hispanic = 0.037 | | | | | | |
| N1N | Not Hispanic, White | 0.389 | 0.389 | 9.724 | 0.000 | 9.949 | 0.000 |
| N4N | Not Hispanic, Asian | 0.269 | 0.659 | 0.470 | 0.054 | 0.191 | 0.009 |
| N2N | Not Hispanic, Black | 0.195 | 0.853 | 0.028 | -0.159 | -0.456 | -0.158 |
| N7N | Not Hispanic, 2+ Races (not SOR) | 0.060 | 0.913 | 0.116 | -0.013 | 0.001 | -0.021 |
| N6N | Not Hispanic, Some Other Race | 0.027 | 0.940 | 0.082 | -0.126 | -0.302 | -0.103 |
| H1N | Hispanic, White | 0.016 | 0.956 | -0.082 | -0.128 | -0.113 | -0.131 |
| XXX | All Other Responses | 0.013 | 0.969 | -0.110 | -0.171 | -0.418 | -0.132 |
| N8N | Not Hispanic, 2+ Races (SOR) | 0.011 | 0.980 | 0.028 | -0.015 | -0.168 | -0.017 |
| N5N | Not Hispanic, Native Hawaiian/Pacific Islander | 0.008 | 0.989 | -0.291 | -0.088 | -0.288 | -0.102 |
| H6N | Hispanic, Some Other Race | 0.007 | 0.996 | -0.221 | -0.164 | -0.189 | -0.139 |
| H8N | Hispanic, 2+ Races (SOR) | 0.004 | 1.000 | -0.167 | -0.164 | -0.123 | -0.124 |
| 2P1OFB | Two Parents, One Other Foreign Born, One Born USA \| Sample Size = 237,000 \| Hispanic = 0.031 | | | | | | |
| N1N | Not Hispanic, White | 0.781 | 0.781 | 9.893 | 0.000 | 10.160 | 0.000 |
| N7N | Not Hispanic, 2+ Races | 0.084 | 0.865 | 0.058 | -0.012 | -0.057 | -0.012 |
| N2N | Not Hispanic, Black | 0.046 | 0.912 | -0.142 | -0.168 | -0.545 | -0.162 |
| N4N | Not Hispanic, Asian | 0.035 | 0.947 | 0.245 | 0.007 | 0.073 | -0.007 |
| H1N | Hispanic, White | 0.017 | 0.964 | -0.081 | -0.110 | -0.239 | -0.138 |
| XXX | All Other Responses | 0.010 | 0.974 | -0.298 | -0.151 | -0.291 | -0.163 |
| N6N | Not Hispanic, Some Other Race | 0.008 | 0.982 | -0.133 | -0.039 | -0.272 | -0.046 |
| N8N | Not Hispanic, 2+ Races (SOR) | 0.008 | 0.990 | -0.122 | -0.051 | -0.300 | -0.021 |
| H6N | Hispanic, Some Other Race | 0.004 | 0.994 | -0.239 | -0.166 | -0.193 | -0.146 |
| N5N | Not Hispanic, Native Hawaiian/Pacific Islander | 0.003 | 0.997 | -0.257 | -0.096 | -0.387 | -0.116 |
| H7N | Hispanic, 2+ Races | 0.003 | 1.000 | -0.029 | -0.145 | -0.106 | -0.130 |
| 2OFB | Two Other Foreign-Born Parents \| Sample Size = 174,000 \| Hispanic = 0.001 | | | | | | |
| N4N | Not Hispanic, Asian | 0.611 | 0.611 | 10.350 | 0.000 | 10.320 | 0.000 |
| N1N | Not Hispanic, White | 0.211 | 0.823 | -0.210 | -0.067 | -0.035 | -0.014 |
| N2N | Not Hispanic, Black | 0.091 | 0.914 | -0.312 | -0.237 | -0.487 | -0.203 |
| N7N | Not Hispanic, 2+ Races | 0.036 | 0.950 | -0.171 | -0.040 | -0.104 | -0.015 |
| N6N | Not Hispanic, Some Other Race | 0.023 | 0.973 | -0.291 | -0.159 | -0.356 | -0.101 |
| N5N | Not Hispanic Pacific Islander | 0.010 | 0.983 | -0.863 | -0.175 | -0.572 | -0.105 |
| N8N | Not Hispanic, 2+ Races (SOR) | 0.010 | 0.993 | -0.251 | -0.067 | -0.143 | -0.027 |
| XXX | All Other Responses | 0.007 | 1.000 | -0.453 | -0.229 | -0.327 | -0.178 |
| 1PUS | Single USA Born Parent \| Sample Size = 1,889,000 \| Hispanic = 0.073 | | | | | | |
| N1N | Not Hispanic, White | 0.675 | 0.675 | 9.546 | 0.000 | 9.860 | 0.000 |
| N2N | Not Hispanic, Black | 0.219 | 0.894 | -0.253 | -0.158 | -0.733 | -0.137 |
| H1N | Hispanic, White | 0.043 | 0.937 | -0.099 | -0.125 | -0.151 | -0.120 |
| H6N | Hispanic, Some Other Race | 0.018 | 0.955 | -0.235 | -0.159 | -0.249 | -0.153 |
| N7N | Not Hispanic, 2+ Races | 0.013 | 0.968 | -0.184 | -0.045 | -0.303 | -0.044 |
| N3N | Not Hispanic, American Indian/Alaska Native | 0.011 | 0.979 | -0.574 | -0.082 | -0.729 | -0.064 |
| XXX | All Other Responses | 0.008 | 0.988 | -0.164 | -0.105 | -0.275 | -0.083 |
| N6N | Not Hispanic, Some Other Race | 0.005 | 0.992 | -0.340 | -0.074 | -0.488 | -0.047 |
| H8N | Hispanic, 2+ Races (SOR) | 0.004 | 0.997 | -0.215 | -0.164 | -0.301 | -0.139 |
| H2N | Hispanic, Black | 0.003 | 1.000 | -0.190 | -0.250 | -0.540 | -0.214 |
| 2US | Two USA Born Parents \| Sample Size = 4,374,000 \| Hispanic = 0.037 | | | | | | |
| N1N | Not Hispanic, White | 0.882 | 0.882 | 9.836 | 0.000 | 10.180 | 0.000 |
| N2N | Not Hispanic, Black | 0.060 | 0.942 | -0.218 | -0.185 | -0.645 | -0.162 |
| H1N | Hispanic, White | 0.025 | 0.968 | -0.162 | -0.138 | -0.201 | -0.141 |
| N7N | Not Hispanic, 2+ Races | 0.008 | 0.976 | -0.226 | -0.049 | -0.287 | -0.056 |
| XXX | All Other Responses | 0.008 | 0.984 | -0.199 | -0.132 | -0.289 | -0.122 |
| H6N | Hispanic, Some Other Race | 0.007 | 0.991 | -0.313 | -0.179 | -0.287 | -0.178 |
| N3N | Not Hispanic, American Indian/Alaska Native | 0.006 | 0.997 | -0.572 | -0.093 | -0.612 | -0.078 |
| N6N | Not Hispanic, Some Other Race | 0.003 | 1.000 | -0.299 | -0.055 | -0.344 | -0.031 |
| 2PDFB | Two Parents, Each Parent in a Different Foreign-Born Group \| Sample Size = 12,000 \| Hispanic = 0.483 \| Census Detail/SSA POB Match = 0.45 | | | | | | |
| N1N | Not Hispanic, White | 0.331 | 0.331 | 9.955 | 0.000 | 10.060 | 0.000 |
| H1P | Hispanic, White, Census Detail/SSA POB Match | 0.220 | 0.551 | 0.061 | -0.157 | 0.067 | -0.127 |
| XXX | All Other Responses | 0.102 | 0.653 | 0.050 | -0.149 | -0.089 | -0.110 |
| H6P | Hispanic, Some Other Race, Census Detail/SSA POB Match | 0.093 | 0.746 | -0.147 | -0.218 | -0.085 | -0.209 |
| N6N | Not Hispanic, Some Other Race | 0.055 | 0.801 | -0.016 | -0.038 | 0.014 | -0.042 |
| H8P | Hispanic, 2+ Races (SOR), Census Detail/SSA POB Match | 0.051 | 0.852 | -0.014 | -0.164 | -0.079 | -0.146 |
| N4N | Not Hispanic, Asian | 0.047 | 0.898 | 0.392 | -0.004 | 0.275 | 0.025 |
| H4P | Not Hispanic, Asian, Census Detail/SSA POB Match | 0.038 | 0.936 | 0.143 | -0.195 | 0.124 | -0.150 |
| N2N | Not Hispanic, Black | 0.034 | 0.970 | 0.157 | -0.194 | -0.258 | -0.231 |
| H1N | Hispanic, White | 0.030 | 1.000 | 0.092 | -0.184 | -0.152 | -0.184 |

Notes: This table uses Sample 2, which consists of 7.196 million native born workers age 25 to 28 (inclusive) in 2010. All workers must have at least one year with positive earnings during the six year period from 2010 to 2015. Zer- earning years are included in the earnings measures. The worker and their parent's POB are determined using the Census Numident. The parent's are identified using the family structure information on the 2000 Census. The race measure consists of eight categories collated from the 2010 Census: White, Black, American Indian/Alaska Native (AIAN), Asian, Native Hawaiian/Pacific Islander, Some Other Race (SOR), 2+ Races (not SOR), and 2+ Races (SOR). The ethnicity variable is binary; Hispanic/Latino or Not Hispanic/Latino. A Census Detail/SSA POB Match occurs when the 2010 Census respondent reported their Census Numident POB on the 2010 Race or Ethnicity questions using the detailed coding (the detailed Decennial POB coding is not available for 8 MENA countries as shown in Table 3). For both females and males we report four measures of mean log real long-term earnings. The first row of the Mean Log Earnings column for each POB group shows the mean log earnings of the largest Ethnicity-Race-Census Detail/SSA POB Match category, while the other rows report the difference in mean log earnings relative to the largest category. The next three columns show adjusted earnings from linear regression models controlling for various person and job characteristics. The PC model includes controls for education and quadratic functions of age and years in the U.S. in 2010. The PCW model adds controls for the dominant industry/region and the number of full and partial years worked. The PCWH model adds a quadratic function of average annual hours worked. Standard errors have not been released yet for this table. Some sub-groups have fewer than 1,000 persons. Even for small sub-groups, standard errors for adjusted log earnings differentials rarely exceed 0.025 log points.